\documentclass{aa}  

\usepackage{graphicx}
\usepackage{txfonts}
\begin{document} 

\title{eROSITA X-ray scan of the $\eta$ Chamaeleontis cluster}
\subtitle{Member study and search for dispersed low-mass stars}

\author{J. Robrade
\and S. Czesla
\and S. Freund
\and J.H.M.M. Schmitt
\and P.C. Schneider
}
\institute{Hamburger Sternwarte, University of Hamburg, Gojenbergsweg 112, D-21029 Hamburg, Germany\\
\email{jrobrade@hs.uni-hamburg.de}
}
\date{Received 19 April 2021; accepted 25 May 2021}

% \abstract{}{}{}{}{} 
% 5 {} token are mandatory
\abstract
{The nearby young open cluster $\eta$~Chamaeleontis has been observed by eROSITA/SRG during its CalPV phase for 150~ks. The extended ROentgen Survey with an Imaging Telescope Array (eROSITA) data were taken in the field-scan mode, an observing mode of Spectrum-Roentgen-Gamma (SRG) that follows a rectangular grid-like pattern, here covering a 5\,x\,5~deg field with an exposure depth of about 5~ks.}
{The $\eta$~Cha cluster with an age of about 8 Myr is a key target for investigating the evolution of young stars, and we aim to study the known members in X-rays. Additionally, we search for potential new members of the anticipated dispersed low-mass cluster population in a sensitive wide-field X-ray observation.}
{Using eROSITA X-ray data, we studied the $\eta$~Cha region. Detected sources were identified by cross-matching X-ray sources with Gaia and 2MASS, and young stars were identified by their X-ray activity, the position in the color-magnitude diagram, and by their astrometric and kinematic properties. X-ray-luminosities, light curves, and spectra of cluster members were obtained and compared with previous X-ray data. Literature results of other member searches were used to verify our new member candidates in the observed field.}
{We determine X-ray properties of virtually all known $\eta$~Cha members and identify five additional stellar systems that show basically identical characteristics, but they are more dispersed. Four of them were previously proposed as potential members; this status is supported by our X-ray study. Based on their spatial distribution, further members are expected beyond the sky region we surveyed. The identified stellar systems very likely belong to the ejected halo population, which brings the total number of $\eta$~Cha cluster members to at least 23.
}
{Sensitive X-ray surveys are best suited to identifying active stars, and the combination of the ongoing eROSITA all-sky survey with Gaia measurements provides an unprecedented opportunity to study the nearby, young stellar population.
}


\maketitle
%
%-------------------------------------------------------------------

\section{Introduction}

The $\eta$~Chamaeleontis ($\eta$~Cha) cluster is a nearby, young open cluster and the first that was discovered principally through X-ray studies by \cite{mam99}. In a ROSAT High Resolution Imager (HRI) pointing, they discovered 12 X-ray sources that were found to be associated with prominent young stellar counterparts. 
The $\eta$~Cha cluster is associated with the Oph-Sco-Cen association (OSCA), which is a large multiple-epoch star formation region. 
The cluster is spatially offset from the main OSCA group. Despite its youth, there is virtually no sign of intervening cloud material. 
With a distance of 97~pc and an age of 8~Myr \citep{mam00}, $\eta$~Cha is one of a group of young clusters that is nearest to the Sun.
The age of the cluster has been re-estimated by multiple authors, and the derived population ages range from $6^{+2}_{-1}$~Myr \citep{luh04} to $11 \pm 3$~Myr \citep{bell15}.
The cluster members are prime targets for studying stellar pre-main-sequence evolutionary processes, accretion properties, disk dispersal, and planet formation, and they have been frequently observed by major ground- and space-based facilities.

The $\eta$~Cha cluster is compact, with a radius of 1~pc. It contains three intermediate-mass stars: the central, eponymous B8V star $\eta$~Cha itself, the early-A star HD~75505, and the eclipsing A-star binary RS~Cha. Additionally, various young stellar objects (YSOs) in the low- to solar-mass range are associated to the cluster and 13 members were identified in the initial study. Furthermore, the $\eta$~Cha cluster exhibits a high disk fraction of about 50\,\%. It also contains stars in various evolutionary stages with different accretion phases and disk morphologies, that is, classical T~Tauri stars (CTTS; SED class II, flat), transition objects (TO), and weak-lined T~Tauri stars (WTTS; SED class III). The diversity of the $\eta$~Cha disks has been noted repeatedly, for instance, in observations with Spitzer/IRAC \citep{meg05}, Spitzer/IRS \citep{sic09}, or Herschel-PACS \citep{riv15}. Ongoing planet formation has been suggested based on studies of protoplanetary disks around several of the $\eta$~Cha stars. A recent campaign with VLT/X-Shooter studies stellar parameter and accretion properties for all low-mass stars of the cluster \citep{rug18}.

\cite{mam00} have noted a deficit of low-mass stars in $\eta$~Cha, indicating that the true extent of the cluster reaches well beyond the FOV of ROSAT. These authors presented cluster evaporation models that suggest a halo of low-mass members within a few degrees of the $\eta$~Cha cluster core. The search for these young stars continued predominantly in the optical. Using accretion (H$\alpha$) and stellar youth (Li) tracers as well as kinematic properties, additional members belonging to the $\eta$~Cha cluster have been identified \citep[see, e.g.,][]{law02, luh04, lyo04, song04}. The acronym 'RECX' is used for all these members, although not all are ROSAT sources. 
Eighteen 'classical' members were established in these early studies. Their properties are summarized in Table.~\ref{mem}.

Based on this configuration, N-body simulations by \cite{mor07} indicate that a very compact formation scenario can explain the apparently peculiar mass distribution of the $\eta$~Cha cluster. Alternative explanations such as an abnormal initial mass function (IMF) are not required; the missing members were just ejected during the dynamical evolution of the cluster. Thus a diffuse halo of missing cluster members should exist. The search for dispersed members is ongoing, and further lithium-rich low-mass stars were discovered by \cite{mur10} in 5.5$^\circ$ wide search regions around $\eta$~Cha. The authors identified several probable or possible new members based on kinematics and lithium diagnostics. Many of them are also included in our X-ray study.

In X-rays, the core of the $\eta$~Cha cluster was subsequently studied in a 50~ks pointing by {\it XMM-Newton} \citep{lop10}.
Further observations include a 10~ks {\it Chandra} ACIS-I pointing from 2004 and a 40~ks {\it XMM-Newton} pointing centered on the adjacent star EG~Cha in 2011. Both are unpublished, and the archival data were analyzed in the context of this study.
Sufficiently deep X-ray data for a study of the dispersed population surrounding the cloud core, which is required to reliably separate cluster members from unrelated field stars with activity measurements, have been lacking so far because the FOV of existing X-ray facilities is limited. 
With the launch of eROSITA on board the SRG spacecraft, a suitable space observatory is available. A detailed description of the eROSITA instrument is presented in \cite{pred21}.

Here we present first results from the eROSITA observation of the $\eta$~Cha field. The observation provides a sufficiently large area coverage and sensitivity to study its members down to stellar masses in the mid-M dwarf regime. Detected sources are cross-matched with Gaia EDR3 \citep{gaiaedr3} and other datasets to identify and characterize the various stellar populations. 
The $\eta$~Cha cluster will be revisited eight times during the eRASS (eROSITA all-sky survey), which started in December 2019, but with a shallower exposure \citep[eROSITA Science Book, ][]{mer12}. An analysis of the complete source catalog of the field and obtained survey data will be presented in another publication. 

Our paper is structured as follows. In Sect.~\ref{obs} we describe the dataset, analysis methods, and the created X-ray source catalog. In Sect.~\ref{res} we present our results for the $\eta$~Cha cluster study subdivided into several topics, and in Sects.~\ref{dis} and \ref{sum} we discuss and summarize our findings.

\begin{table*}[t]
\caption{\label{mem}List of known $\eta$~Cha members.}
\begin{center}
\begin{tabular}{rrrrrrrc}\hline\hline\\[-3.1mm] 
RECX & Name & G & BP-RP & Sp. type(s) &  H$\alpha$ & SED class & remarks \\\hline\\[-3mm]
1 & EG Cha                & 10.04 &  1.57  & K7+M0      &  -1       &  III  &   M99 \\
2 & $\eta$ Cha            & 5.44  & -0.13  & B8         &           &  debris & M99 \\
3 & EH Cha                & 13.09 &  2.54  & M3         &  -2/-3    &  TO   &   M99, L10\\
4 & EI Cha                & 11.91 &  2.06  & M1         &  -2/-4    &  TO   &   M99, L10\\
5 & EK Cha                & 13.72 &  2.86  & M4         &  -14/-35  &  TO   &   M99, L10\\
6 & EL Cha                & 13.00 &  2.50  & M3         &  -4/-5    &  III  &   M99, L10\\
7 & EM Cha                & 10.38 &  1.57  & K7+M1      &  0/-1     &  III  &   M99, L10\\
8 & RS Cha                & 6.01  &  0.33  & A7+A8+?    &           &  III  &   M99, L10\\
9 & EN Cha                & 13.80 &  3.13  & M4.5+M4.5  &  -10/-13  &   TO  &   M99, L10\\
10 & EO Cha               & 11.80 &  1.92  & M0         &  -1/-2    &  III  &   M99, L10\\
11 & EP Cha               & 10.67 &  1.54  & K6.5       &  -3/-8    &   II  &   M99, L10\\
12 & EQ Cha               & 12.04 &  2.46  & M3+M3      &  -4/-8    &  III  &   M99, L10\\
13 & HD 75505             & 7.37  &  0.22  & A1         &           &  III  &   \\
14 & ES Cha               & 15.15 &  3.43  & M5         &  -10      &  flat/TO &   L10 \\
15 & ET Cha               & 13.34 &  2.33  & M3+?       &  -90/-95  &  II   &  L10  \\
16 & 2M J08440914-7833457 & 16.22 &  3.32  & M5.5       &  -58/-102 &  flat &   \\
17 & 2M J08385150-7916136 & 14.78 &  3.53  & M5+M5      &  -10/-12  &  III  &   \\
18 & 2M J08361072-7908183 & 15.42 &  3.76  & M5.5+M5.5  &  -8/-13   &  III  &   \\\hline
\end{tabular}
\end{center}
\tablefoot{H$\alpha$ EW from \cite{law04, song04,rug18}, G and BP-RP magnitudes from Gaia EDR3, SED class from \cite{sic09}, remarks: M99: X-ray source (RECX) in \cite{mam99}, L10: X-ray source in \cite{lop10}.}
\end{table*}

\section{Observation and data analysis}
\label{obs}

\begin{figure}
\includegraphics[width=88mm]{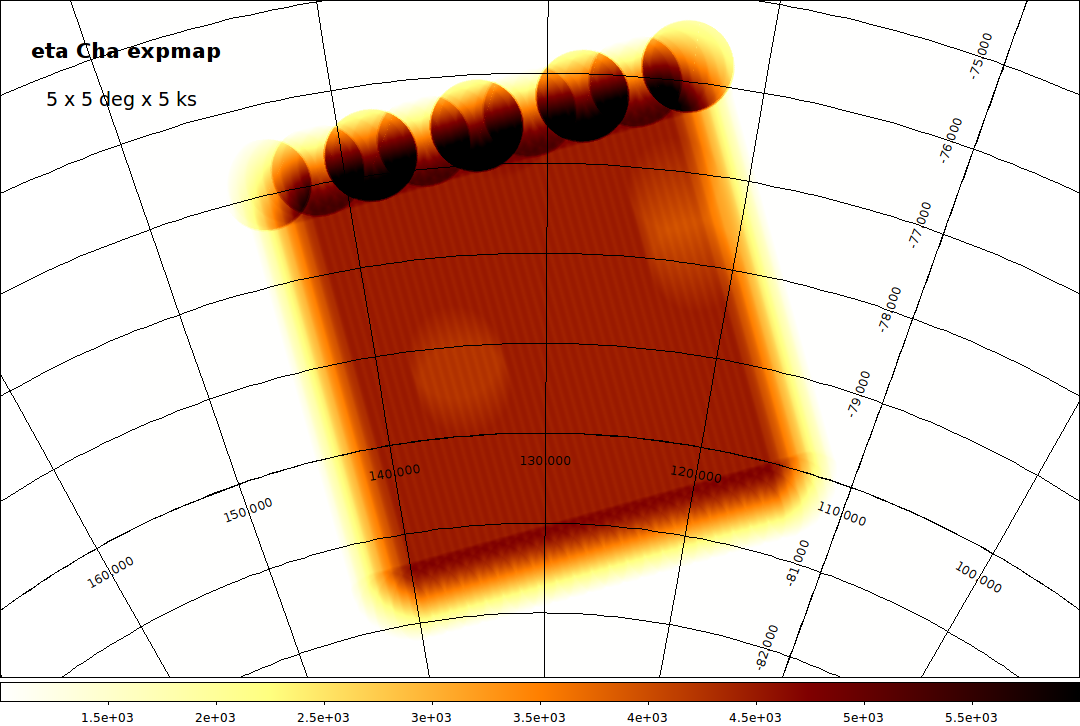}
\includegraphics[width=88mm]{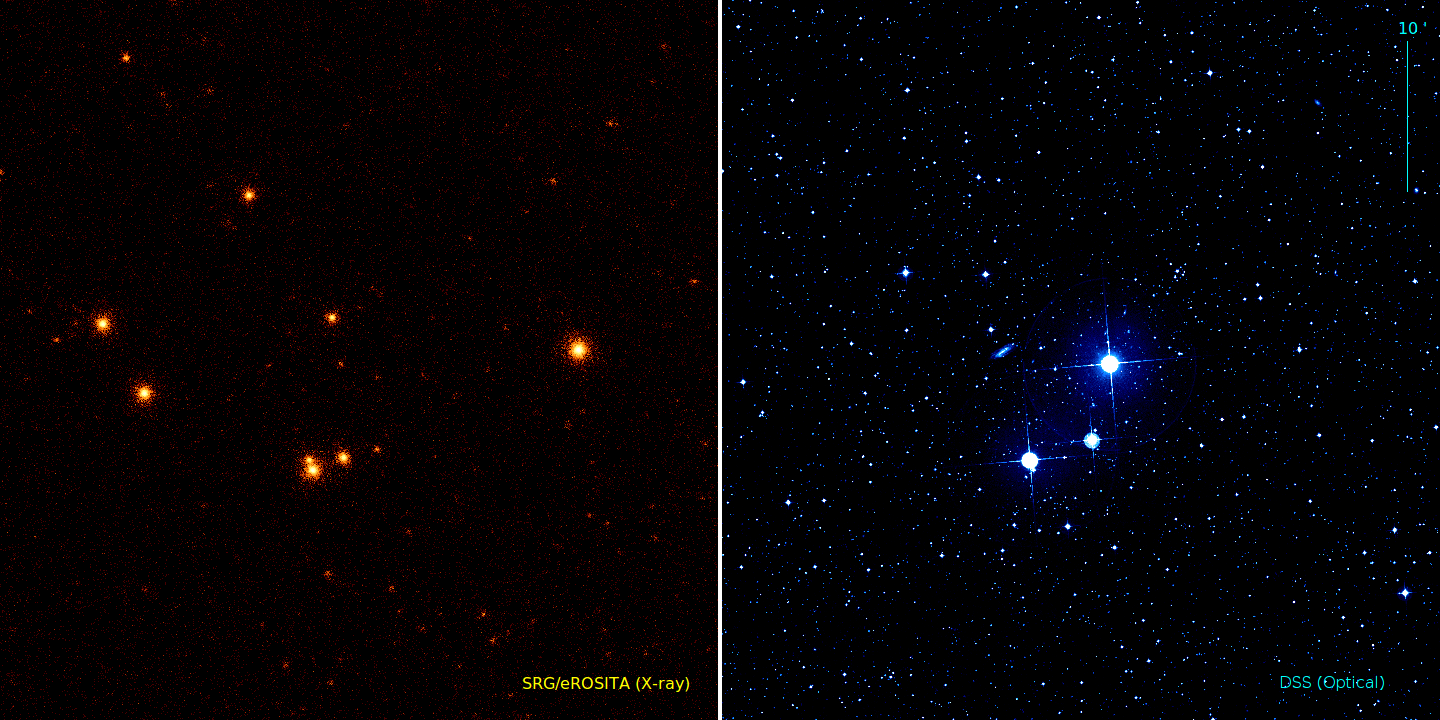}
\includegraphics[width=88mm]{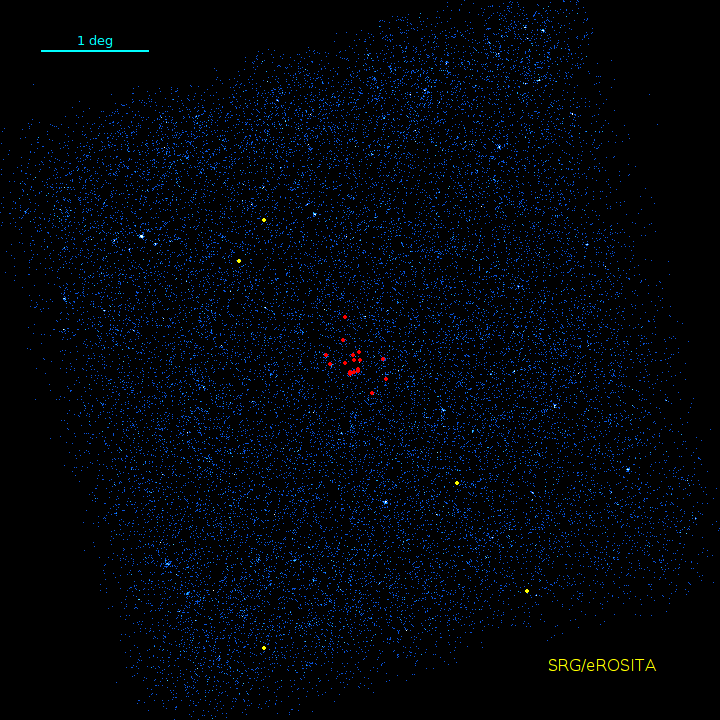}
\caption{\label{ximg}Overview of the eROSITA $\eta$~Cha field scan observation. X-ray exposure map ({\it top}) with linear scaling to facilitate showing the scanning direction and spacecraft maneuver holding points. The color scale denotes exposure in [s]. A comparison of the X-ray and optical images of the cluster core showing the same sky region ({\it middle}), and an X-ray image of the $\eta$~Cha field ({\it bottom}) in which known (red) and new (yellow) cluster members are marked. The field contains nearly 8000 X-ray sources.}
\end{figure}

The $\eta$~Cha field was chosen as a performance verification (PV) and early-science observation of eROSITA. It was carried out in the field-scan mode, a unique X-ray observing mode specific to SRG, in which a sky region is observed in a rectangular grid-like pattern \citep{pred21}. In this mode, each source is repeatedly scanned by the sweeping telescope, and photons are collected during multiple passes through the field of view (FOV) of the instrument at different off-axis angles. The designed field template has an extent of 5 x 5 deg and consists of 25 pattern passes, that is, 50 scan stripes in total with a separation of 0.1~deg cover the field. As eROSITA has a FOV diameter of about 1.03~deg, a 0.5~deg wide region in which the exposure depth fades out surrounds the main field. The observation (OBSID 300004) was performed from 16.11.2019 23:30 - 18.11.2019 18:15 (UTC), resulting in a total observing time of about 154~ks, including overhead. All seven eROSITA telescopes were operational, and the 'FILTER' setting was used to reduce contamination by optical light. The eROSITA instrument covers the 0.2\,--\,10.0~keV energy range and achieves an angular resolution of about 26\,'' half-energy width (HEW) in field-scan mode. The processed data have an unvignetted exposure depth of about 4.5~ks, and the field center is the B8V star $\eta$~Cha at RA/DEC 130.33, -78.96 deg, corresponding to a galactic longitude/latitude of 292.40, -21.65 deg. An overview of the eROSITA $\eta$~Cha field is shown in Fig.~\ref{ximg}.

The exposure depth provides a flux sensitivity of roughly $5 \times 10^{-15}$~erg\,s$^{-1}$\,cm$^{-2}$ in the 0.2\,--\,5.0~keV band for active coronal sources, corresponding to an X-ray brightness limit of $L_{\rm X} = 5\times 10^{27}$~erg\,s$^{-1}$ at a cluster distance of 97~pc. The sensitivity is sufficient to detect young, active stars down to the mid-M~dwarf regime close to the end of the main sequence, assuming their X-ray emission is at or close to the saturated activity level, that is, $\log L_{\rm X}/L_{\rm bol} \approx -3$.
At $M_{*} \approx 0.1 M_{\odot}$ , we expect to detect 10~Myr old stars down to activity levels of $\log L_{\rm X}/L_{\rm bol} = -4$ and 100~Myr old stars down to $\log L_{\rm X}/L_{\rm bol} = -3.5$. Thereby, the eROSITA scan provides a virtually complete sample of the active cluster population at stellar masses of $M_{*} \gtrsim 0.1 M_{\odot}$ in the surveyed region.

\subsection{Data processing and analysis}
\label{xdat}

The eROSITA data were processed with the standard pipeline in version c001, and the merged events from all seven telescopes were taken as input. The eROSITA science analysis software in version eSASSusers\_201009 was used to produce the X-ray source catalog and to generate data products such as spectra and light curves. A detailed description of the eROSITA software, catalog creation procedures, and standard data products is presented in \cite{esass}.

The preparatory post-processing of the initial event file that was specifically performed here includes a small boresight correction and the removal of identified data artifacts caused by two flickering pixels.
The astrometric correction was determined by cross-matching the bright X-ray sources (\texttt{DET\_LIKE~$> 50$}) with Gaia EDR3 sources. A mean positional offset in RA/DEC of 0.0044/-0.0010 deg was found, and the X-ray event positions were corrected accordingly. This modified event file is used in our subsequent analysis.

Adapted to the eROSITA energy bands, specifically 0.2\,--\,0.6 (soft), 0.6\,--\,2.3 (medium) and 2.3\,--\,5.0~keV (hard), the primary source detection runs were performed in the 0.2\,--\,2.3~keV energy range, that is, soft and medium band combined. This selection is most sensitive for typical stellar targets, and the vignetted exposure in this energy band is 2.4~ks averaged over the field. Source detection was made in a chain of several subtasks. In the initial \texttt{erbox} step, its parameters were varied to achieve an optimized performance in the detection runs for our data. These were then used to generate the input for the creation of the main catalog. From the respective \texttt{erbox} output, we generated the source catalogs with the \texttt{ermldet} and \texttt{catprep} tools. In Sect.~\ref{xcat} we describe specific used methods as well as the basic statistics of the final X-ray catalog.

To study individual sources in greater detail, photons were extracted from circular regions around the respective X-ray position; the background was taken from large annuli or nearby regions. For all sources, vignetting, instrument performance, etc. was taken into account. Light curves were corrected for these effects and correspond to the full instrument on-axis equivalent values, as is common for most eROSITA data products. Spectra make use of all source photons. The light curves and images refer to the 0.2\,--\,2.3~keV energy band, as does the main catalog.

Spectral analysis was carried out with XSPEC V12.9 \citep{xspec}, and absorbed multi-temperature APEC \citep{apec} models were used to fit the X-ray spectra. The metallicity of the thermal plasma is poorly constrained by our data and was set to 0.3 solar abundances, a typical value derived from X-ray spectra of YSOs. Spectra were rebinned to a minimum of three counts per bin, and the model optimization used the `cstat' algorithm, which is applicable for Poisson-distributed data. This approach allowed us to treat the data from the different sources independent of their flux in an identical fashion without losing a significant spectral resolution. The $1 \sigma$ errors denote the one sigma confidence range. We tested different binning and modeling approaches and obtained overall consistent results. A comparative example is given in Sect.\ref{xhist} for the X-ray bright WTTS EG~Cha (RECX1). 
From the applied spectral models we determined an average energy conversion factor (ECF) of $0.9 \times 10^{-12}$~erg\,s$^{-1}$ per cts\,s$^{-1}$ that was used to calculate fluxes from measured count rates for all X-ray sources in the catalog. This value reproduces the source flux typically to within 10\,\% when compared to spectral fit results obtained for active stars.

\subsection{X-ray source catalog}
\label{xcat}

To produce the main 0.2\,--\,2.3~keV X-ray source catalog, three detection pipeline parameter settings were used in the initial \texttt{erbox} step, which is a sliding-box detection algorithm that provides the input for the catalog creation. The first setting uses an iterative method, where \texttt{erbox} is run twice. First on the image, to create an initial source list ('local mode'), then a background map with the source list as input, followed by a repeated boxtection using image plus background map ('map mode'). This is the primary method, and its output is fed into \texttt{ermldet} to determine the source parameter. We used a detection likelihood threshold of \texttt{DET\_LIKE} $\ge 6$ . This set is complemented with highly significant sources (\texttt{DET\_LIKE} $>10/20$ for point-like/extended) that are exclusively detected in the local mode run with and without image rebinning. We used the better sensitivity that in specific cases is achieved by this method, for example, in crowded environments, whereas the high threshold ensures a quality cut on the added sources.
The sophisticatedly combined main catalog achieves a high degree of completeness and simultaneously reduces the inclusion of spurious detections.

\begin{figure}[t]
\includegraphics[width=90mm]{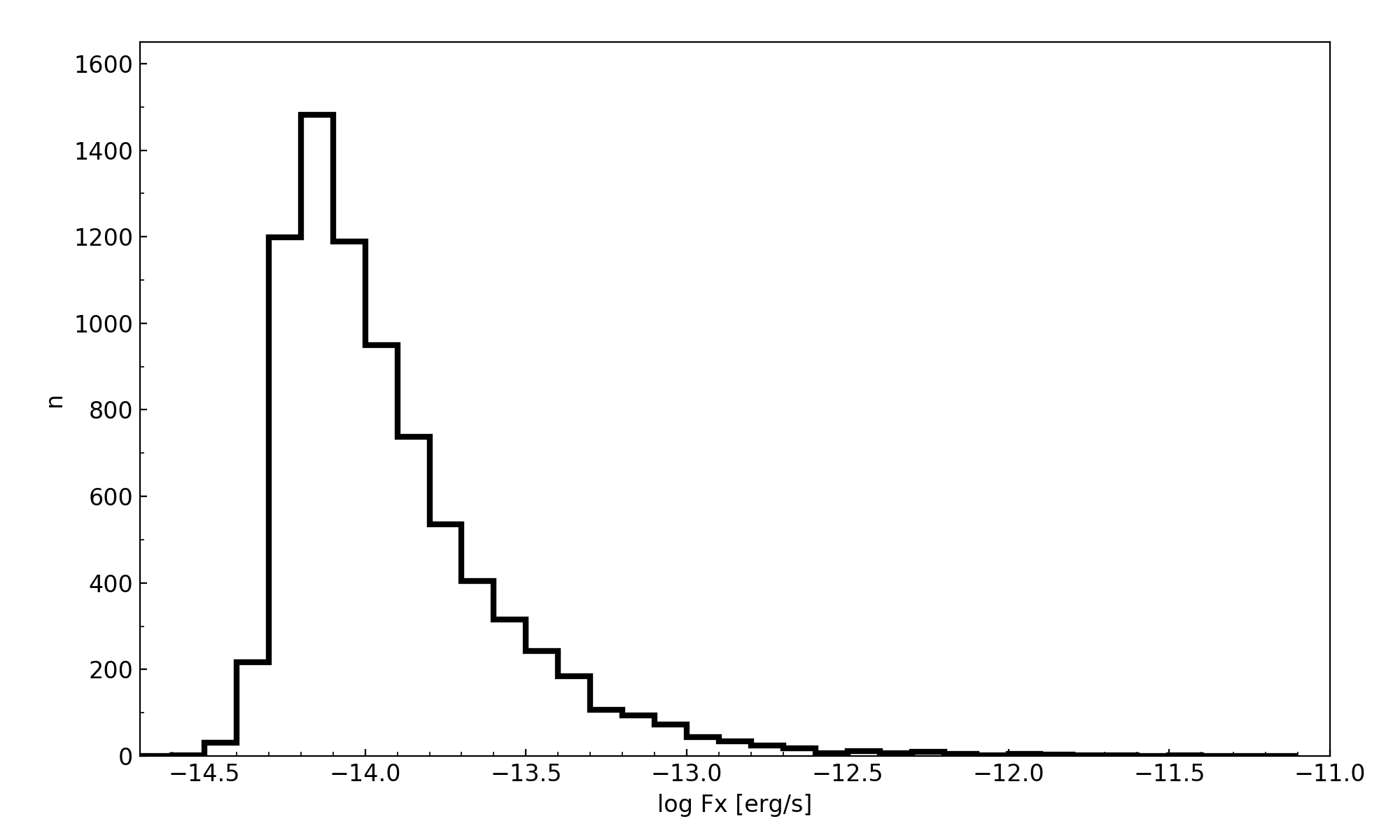}
\caption{\label{fxdistr}Distribution of X-ray fluxes of detected sources in the $\eta$~Cha field in the 0.2\,--\,2.3~keV energy range.}
\end{figure}

A supplementary hard 2.3\,--\,5.0~keV catalog, created solely with the primary \texttt{erbox} method, but with a higher threshold of \texttt{DET\_LIKE}~$\ge 12$, is provided in addition. In eROSITA nomenclature, the ECFs [cts\,cm$^{2}$\,erg$^{-1}$] used to convert count rates into fluxes are 1.1e12 (main) and 1.4e11 (hard). A description of the X-ray catalogs is given in Appendix~\ref{app}.

The final 0.2\,--\,2.3~keV energy band source catalog ('main') of the $\eta$~Cha field observation contains 7933 X-ray sources. One hundred and four of these are classified as extended (\texttt{EXT\_LIKE} $\ge 8$). Of the 7829 point-sources, 5233 have \texttt{DET\_LIKE} $\ge 10,$ and 2050 sources have \texttt{DET\_LIKE} $\ge 30$. The X-ray flux distribution of the detected sources is shown in Fig. \ref{fxdistr}. The source distribution starts to drop out at fluxes below $6 \times 10^{-15}$~erg\,s$^{-1}$\,cm$^{-2}$. The faintest detected sources have a flux of $f_{\rm X} \approx 3 \times 10^{-15}$~erg\,s$^{-1}$\,cm$^{-2}$.

\section{Results}
\label{res}
We report the results we obtained from the X-ray observation of the $\eta$~Cha cluster. We describe the identification of the relevant X-ray sources, study the detected stellar population, and search for new cluster members and candidates. We then analyze the X-ray brighter $\eta$~Cha stars in more detail and compare our results to previous X-ray observations.

\begin{figure}[t]
\includegraphics[width=90mm]{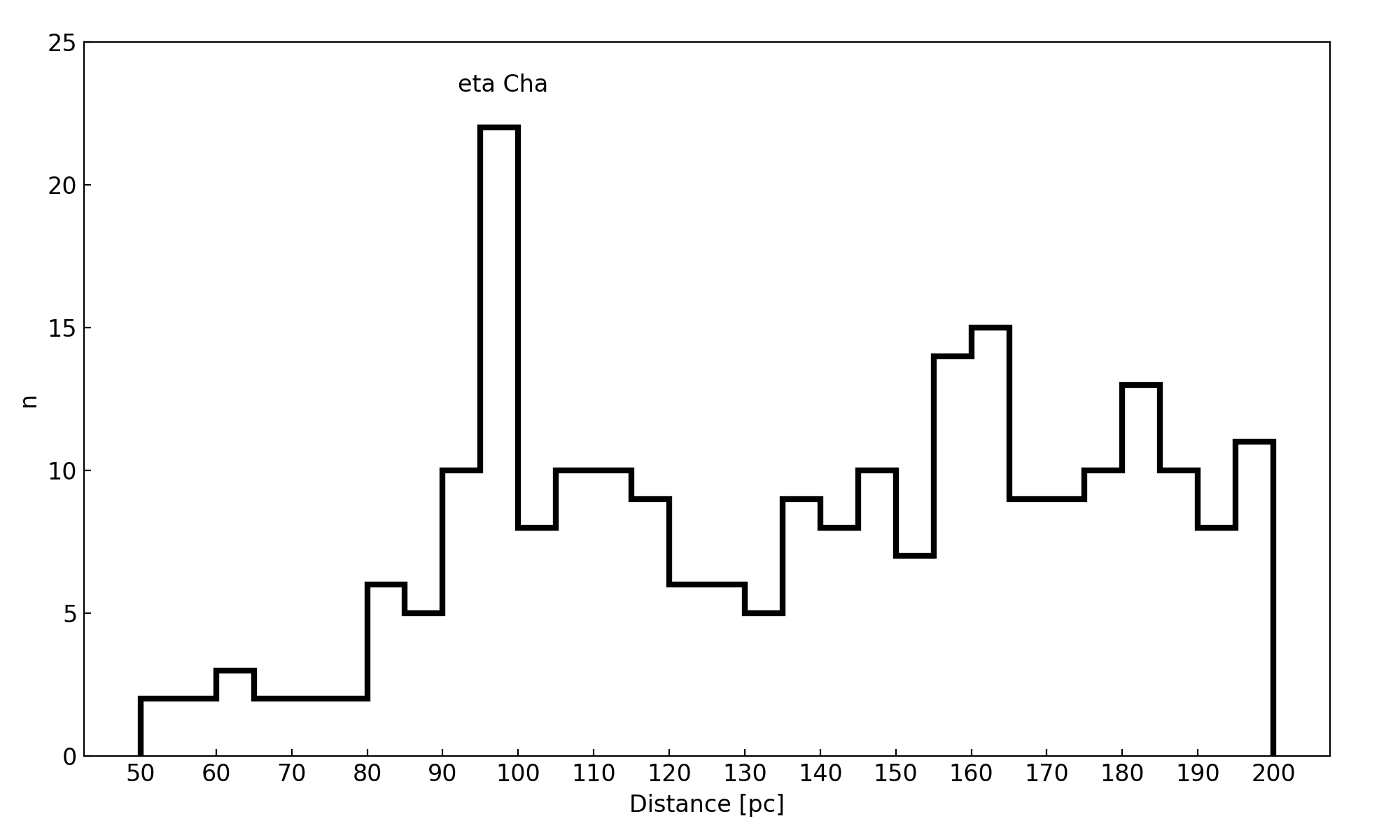}
\caption{\label{xdist}Distances of the X-ray sources in the field that are cross-matched with Gaia EDR3. We show the range 50\,--\,200 pc. The mean $\eta$~Cha cluster distance is labeled.}
\end{figure}

\subsection{Identifications and populations}
\label{ids}

For the identification process, a cross-match of the detected X-ray sources with Gaia EDR3 \citep{gaiaedr3} was performed. Furthermore, 2MASS counterparts \citep{2mass} were added when they lay within 3\arcsec of the epoch 2000 position. As input we used all Gaia sources with a limiting magnitude of G~$\le 20$~mag at their epoch 2020 positions that are located within 10\arcsec of the eROSITA X-ray sources. 
Positional errors are strongly dominated by the X-ray measurement, and for matched sources within 200~pc, for instance, we have a mean positional error of about 3\arcsec. This is comparable to the mean separation between the eROSITA and Gaia position.

\begin{figure*}[ht]
\includegraphics[trim={1.cm 0.5cm 1.cm 1.5cm}, clip, width=180mm]{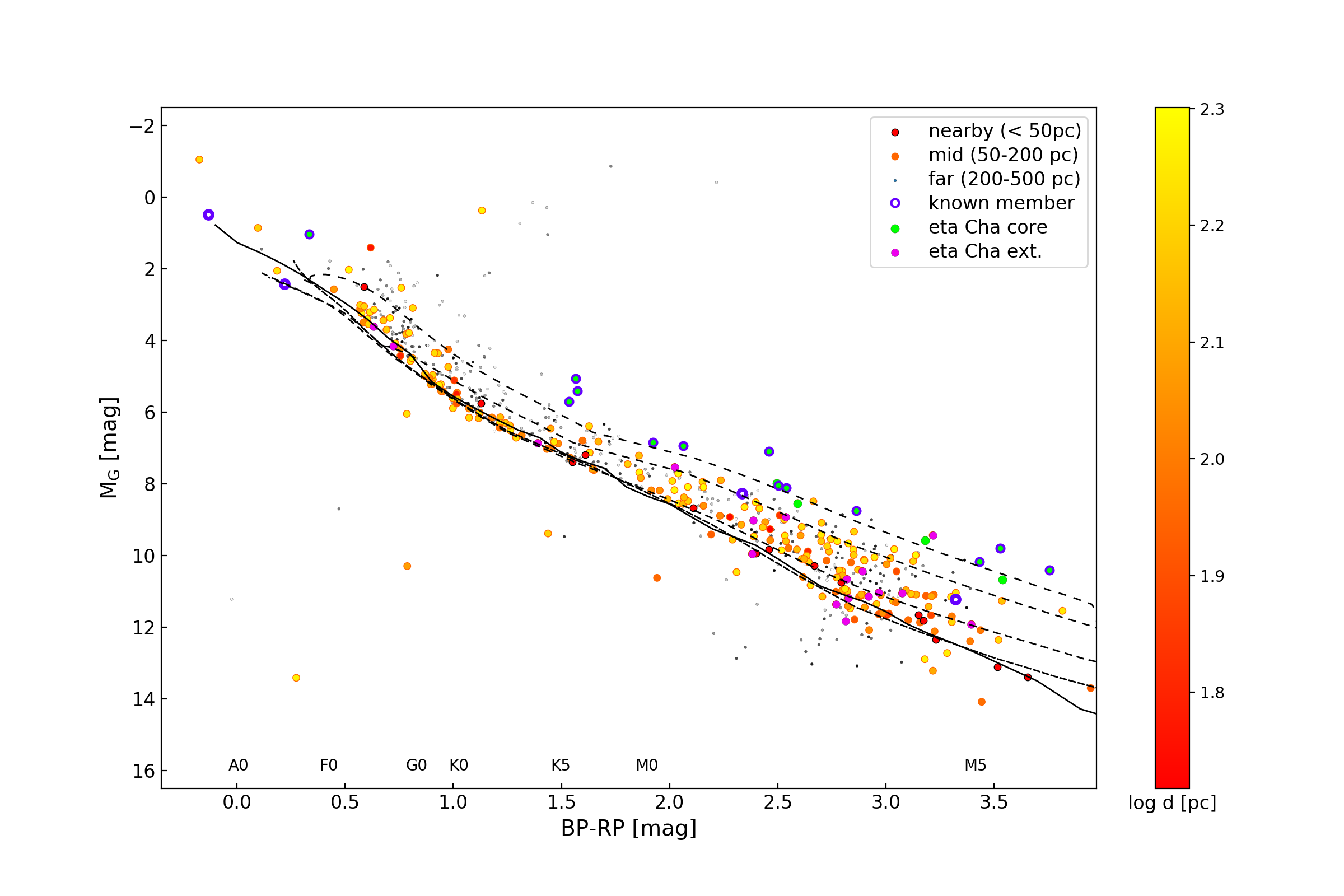}
\caption{\label{xhrd}Gaia CMD of detected stellar X-ray sources, $\eta$~Cha members. Identified core and extended  region stars are highlighted. The color bar represents objects in the 50\,--\,200~ pc distance range, nearby stars are shown as red circles, and more distant stars as gray dots. $\eta$~Cha members and core region members are highlighted. Isochrones are from PARSEC with logAge 7.0, 7.4, 8.0, and 9.0 (dashed line) and MS stars are taken from Mamajek (solid line).}
\end{figure*}

We find about 1000 stellar X-ray sources in the field, which is a fraction of 13\,\% of all point sources. Similar numbers are obtained by machine-learning techniques using the support vector machine (SVM) approach \citep{svm}.
Because we are specifically interested in the nearby stellar component, a step-wise filtering process of the matches is applied in the subsequent analysis. The contamination by spurious identifications was monitored by random sample matching during this process. The contamination is about 1\% and therefore negligible for the sources we discuss.
First, distance cuts were applied to select nearby galactic sources that are potential cluster members or located in the discussed star-forming regions. We specifically focused on stars within 200~pc distance, separated in a nearby zone ($d < 50$~pc) and a middle zone ($50 < d < 200$~pc). 
For sources for which we found multiple possible counterparts, a weighting scheme based on matching distance and source properties such as proximity and brightness was used to identify the most reliable counterpart. In several cases we identified close binaries or multiple systems that were unresolved in X-rays as counterparts. Typically, the primary is given as the contribution from the individual components to the X-ray flux uncertain, in other cases, Gaia also lists only one source.
As eROSITA is prone to optical contamination, brighter stars with $G < 4.5$~mag were verified using their photon energy distribution. Stars that were exclusively detected in the soft band were removed.

To create a color-magnitude diagram (CMD), the Gaia sources are required to have a known distance, that is, parallax and BP-RP color. Gaia EDR3 is not complete in this respect for the targets of interest. A few X-ray sources for which only position and G-band magnitude were available may therefore appear only in the identification list. The Gaia BP-RP color was used to determine approximate spectral type, bolometric correction, and absolute magnitude. We used tabulated values\footnote[1]{A Modern Mean Dwarf Stellar Color and Effective Temperature Sequence, {\it http://www.pas.rochester.edu/$\sim$emamajek}}.

We identify 250 stellar sources at distances up to 200~pc with high confidence. These build the sample that we analyze in greater detail below.
The distance distribution of selected X-ray sources is shown in Fig.~\ref{xdist}. In addition to the $\eta$~Cha cluster at a mean Gaia distance of 99~pc, which is visible in the distribution as a sharp peak, a broader distribution of sources is present in the 90\,--\,120~pc range. At larger distances, further source concentrations are present, particularly at about 160~pc.

A color-magnitude diagram of the detected stars is shown in Fig.~\ref{xhrd}. Isochrones from PARSEC models \citep{parsec} and an MS model from the compilation by E. Mamajek that is largely based on \cite{pec13} are overplotted. In addition to the 200~pc distance sample,
stars found in the far zone ($200 < d < 500$~pc) are plotted for comparison. The fainter grayscale corresponds to more distant objects. Additionally, the known $\eta$~Cha cluster members, including the non-X-ray emitting intermediate-mass stars, are highlighted. Typically, the known low-mass cluster members are located at or above the 10~Myr isochrone in the Gaia CMD, whereas the stellar systems located above this isochrone are typically known binaries or multiples. Exceptions are RECX~15 and RECX~16, the two stars that were also missed in the source detection. They are located below the 10~Myr isochrone, but both are CTTS that possess significant disks, and they are the most strongly accreting systems as measured by the H$\alpha$~EW  in the sample. Thus they are likely also heavily absorbed and reddened.

\begin{figure*}[ht]
\includegraphics[trim={1.cm 0.5cm 1.cm 1.5cm}, clip, width=180mm]{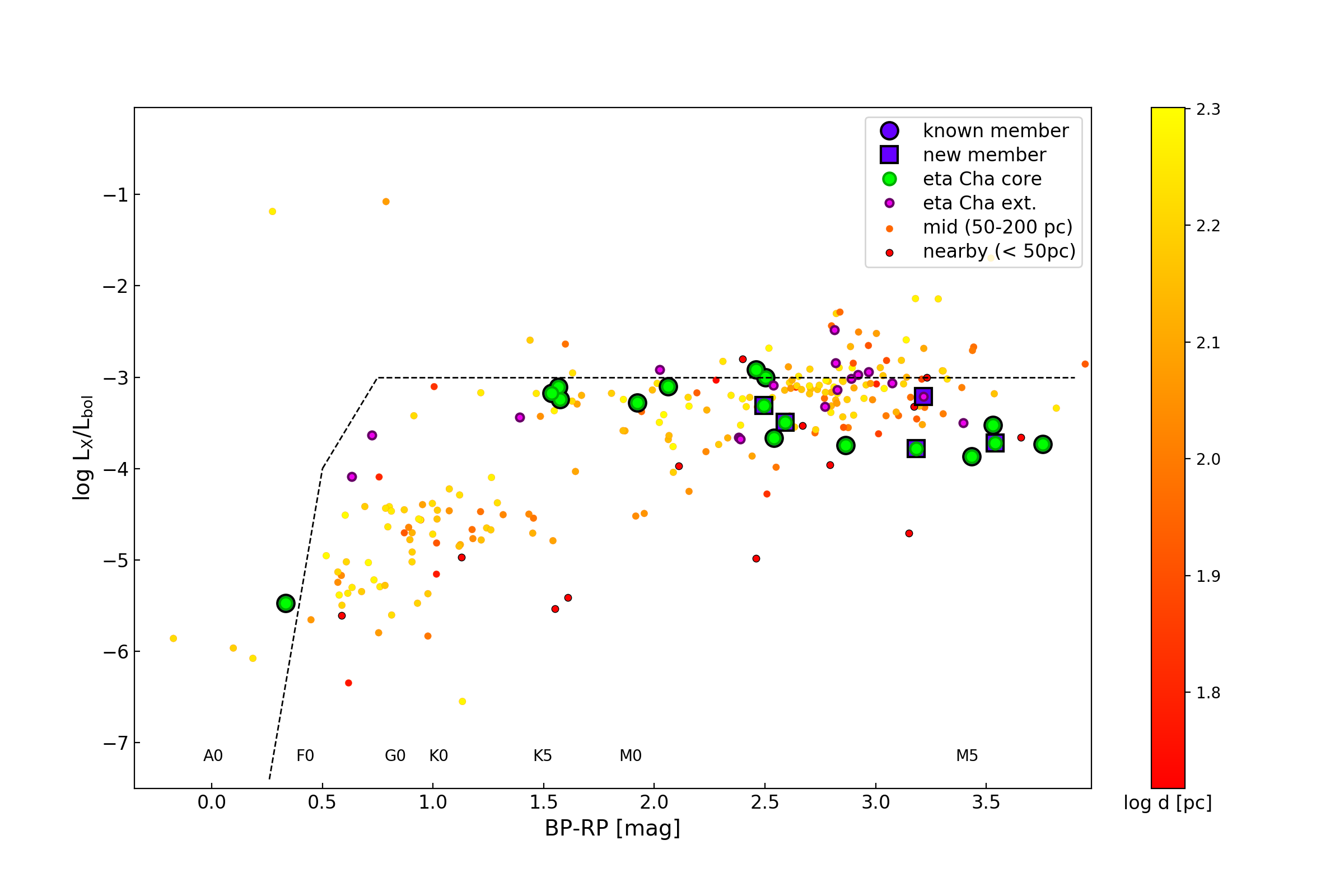}
\caption{\label{xact}X-ray activity level of the detected sources in the $\eta$~Cha field with distances of $\le 200$~pc. The magnetic activity saturation limit is indicated by the dashed line.}
\end{figure*}

We adopted the mean Gaia EDR3 values of known cluster members to define search regions for the ejected members. Specifically, we used a distance $d = 99$~pc and proper motions of $\mu_{\alpha} cos(\delta) = -30$~mas\,yr$^{-1}$ and  $\mu_{\delta} = 27$~mas\,yr$^{-1}$ as average cluster parameter.
In addition, we required the candidates to be X-ray active with the limit at 1 dex below saturation level, that is, with $\log L_{\rm X}/L_{\rm bol} \ge -4$ for late-type stars. 
Combined, these criteria define our initial search regions for new $\eta$~Cha members as follows.
A 'core' region that is defined by $d = 99 \pm 10$ pc, $\mu_{\alpha*} = -30 \pm 5$ mas\,yr$^{-1}$, and $\mu_{\delta} = 27 \pm 5$ mas\,yr$^{-1}$ and an 'extended' region with $d= 99 \pm 30$ pc, $\mu_{\alpha*} = -30 \pm 15$ mas\,yr$^{-1}$, and $\mu_{\delta} = 27 \pm 15$ mas\,yr$^{-1}$.

Our core region includes the kinematic properties and distances of the known cluster members, but already expands the parameter space defined by these stars. Stars ejected during the later cluster evolutionary phase or with moderate speed are expected to be located in this region. Furthermore, an extended search regions is defined by increasing the search parameter by a factor of three to test the detected stars for cluster members that were ejected early or with higher velocities. The stars that fulfill one of these criteria, 19 in the core and 18 in the extended region, are highlighted in Fig.~\ref{xhrd}. All known cluster members in the X-ray catalog indeed satisfy the core criterion.

Magnetically active stars cover a well-defined range of activity levels. The ratio of X-ray to bolometric or optical flux can therefore be used to classify counterparts or identify incorrect associations. Stellar magnetic activity saturates in X-rays at about $\log L_{\rm X}/L_{\rm bol} = -3$ for stellar coronal sources with spectral types F\,--\,M, and the saturation level shows a sharp decline toward hotter stars in the regime of A stars because the outer convective envelope vanishes. The activity level depends on the dynamo efficiency, which is related to stellar rotation and declines by orders of magnitudes with age as a result of magnetic breaking and spin-down to inactive levels at $\log L_{\rm X}/L_{\rm bol} \approx -7$. A comprehensive overview of X-ray properties of magnetically active stars is given in \cite{gue04}, for example.

In Fig.~\ref{xact} we show such an X-ray activity diagram of the detected stars. The known $\eta$~Cha stars and suggested new members are again highlighted. Notably, detected $\eta$~Cha members at spectral types K to early M are found to be about 0.5 dex more active on average than the mid-M~dwarfs in terms of activity level $\log L_{\rm X}/L_{\rm bol}$. While the earlier dwarfs exhibit an average $\log L_{\rm X}/L_{\rm bol} = -3.1$ and thus emit at saturation level, the later dwarfs are overall less active with an average $\log L_{\rm X}/L_{\rm bol} = -3.6$. This is in contrast to the 20\,--\,100~Myr old population found in the vicinity of the cluster, where both groups show an activity level at the saturation limit.

\begin{figure}[t]
\includegraphics[width=90mm]{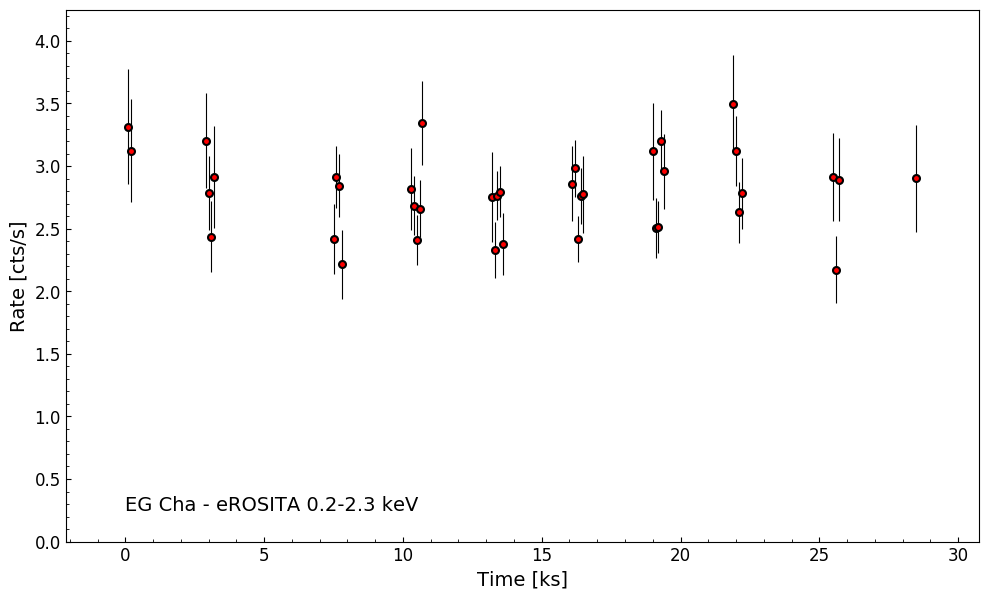}
\includegraphics[width=90mm]{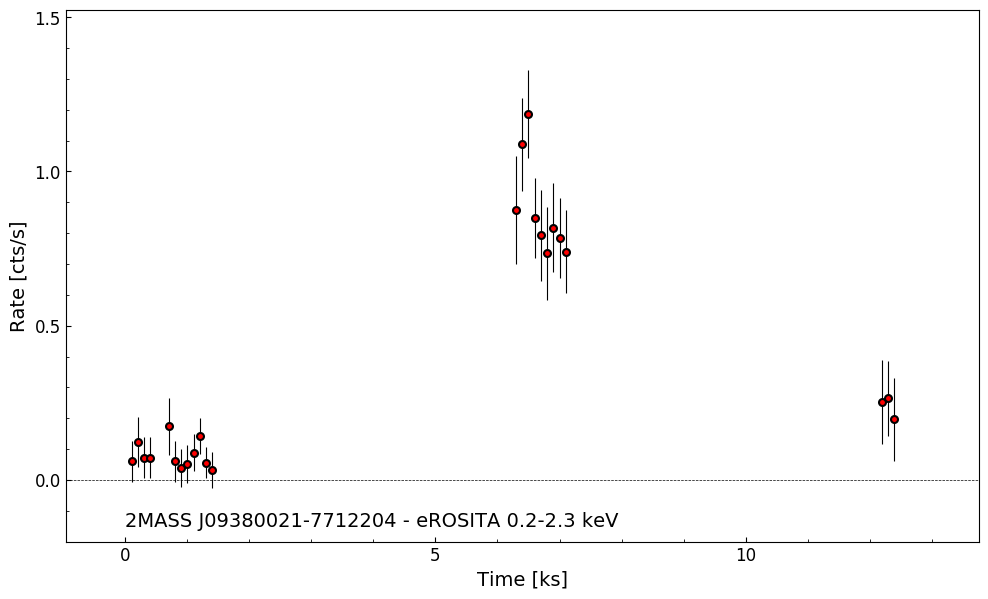}
\caption{\label{xlc} X-ray light curves of EG Cha (RECX1, top) and a flaring M-dwarf located in front of the cluster (bottom). Both are vignetting corrected and have a time binning of 100~s. tzero refers to the respective first time bin.}
\end{figure}

Some stars, particularly M dwarfs, are detected above the canonical saturation limit. These sources are typically observed during a stronger flare or flaring phase that temporarily causes the observed elevated activity level. In Fig.~\ref{xlc} we show an example of such a star, which is detected with an average $\log L_{\rm X}/L_{\rm bol} = -2.3$ during our observation. 

A different class are the highly active blue objects. These are not coronal sources, but are associated with accreting stellar sources. An example in our field is the cataclysmic variable star Z~Cha at a distance of 120~pc with $\log L_{\rm X}/L_{\rm bol} = -1.0$. Another at first look unexpected class are the apparently weakly active blue sources beyond the coronal boundary at spectral types mid A. If they are the true counterpart, these are likely X-ray detections due to an typically unknown low-mass companions of the primary. As discussed in Sect.~\ref{recx}, the $\eta$~Cha cluster member RS~Cha likely also belongs to this category.

\subsection{Young stars and search for new members}
\label{search}

\begin{table*}[t]
\caption{\label{xmem}List of known and new members and candidates.}
\begin{center}
\begin{tabular}{lrrrcrrrrr}\hline\hline\\[-3.1mm]
Target & $L_{\rm X~obs}$ &  log$L_{\rm X}$/$L_{\rm bol}$ & $G$ & $G_{\rm BP}-G_{\rm RP}$ & $d$ & $\mu_{\alpha*}$ / $\mu_{\delta}$ & $\eta_{\rm sep}$ & remarks\\\hline\\[-3mm]
2MASS   & erg\,s$^{-1}$ &  & mag & mag & pc &  mas\,yr$^{-1}$ & pc & \\\hline\\[-3mm]
J08361072-7908183 & 2.2e+28 & -3.73      & 15.4 & 3.76  & 100.9 & -29.1 / 27.9  & 2.6  &  \object{RECX 18}\\  
J08365623-7856454 & 2.8e+30 & -3.11      & 10.0 & 1.57  &  98.9 & -29.6 / 26.9  & 0.7  &  \object{RECX 1} \\
J08385150-7916136 & 5.0e+28 & -3.53      & 14.8 & 3.53  &  99.2 & -28.9 / 27.0  & 1.0  &  \object{RECX 17}\\
J08411947-7857481 &   -     &    -       &  5.4 & -0.13 &  98.3 & -29.4 / 26.8  & 0.0  &  \object{RECX 2} \\
J08413030-7853064 & 1.5e+28 & -3.87      & 15.2 & 3.43  &  98.8 & -30.0 / 28.0  & 0.5  &  \object{RECX 14}\\
J08413703-7903304 & 8.5e+28 & -3.67      & 13.1 & 2.54  &  98.8 & -28.4 / 27.0  & 0.5  &  \object{RECX 3} \\
J08414471-7902531 &   -     &    -       &  7.4 & 0.22  &  98.0 & -30.7 / 27.1  & 0.4  &  \object{RECX 13}\\
J08422372-7904030 & 7.1e+29 & -3.10      & 11.9 & 2.06  &  98.7 & -30.8 / 26.0  & 0.4  &  \object{RECX 4}\\
J08422710-7857479 & 4.7e+28 & -3.75      & 13.7 & 2.86  &  98.5 & -30.2 / 26.9  & 0.2  &  \object{RECX 5}\\
J08423879-7854427 & 4.1e+29 & -3.00      & 13.0 & 2.50  &  98.1 & -29.1 / 27.0  & 0.3  &  \object{RECX 6}\\
J08430723-7904524 & 1.5e+30 & -3.24      & 10.4 & 1.57  &  98.7 & -30.2 / 27.6  & 0.5  &  \object{RECX 7}\\
J08431222-7904123 & 3.8e+29 & -5.47      &  6.0 & 0.33  &  98.6 & -27.3 / 28.2  & 0.4  &  \object{RECX 8}\\
J08431857-7905181 & 2.6e+28 & -4.0       & 13.3 & 2.33  & 103.4 & -27.9 / 29.2  & 5.0  &  \object{RECX 15}\\
J08440914-7833457 &$<$ 4.2e+27 &$<$ -4.0 & 16.2 & 3.32  &  99.6 & -29.8 / 26.3  & 1.4  &  \object{RECX 16}\\
J08441637-7859080 & 7.9e+27 & -4.66      & 13.8 & 3.13  &  @100 &  -- / --      & 1.7  & $G_{\rm phot}$, \object{RECX 9}\\
J08443188-7846311 & 5.0e+29 & -3.28      & 11.8 & 1.92  &  98.2 & -30.3 / 26.9  & 0.5  &  \object{RECX 10}\\
J08470165-7859345 & 1.4e+30 & -3.17      & 10.7 & 1.54  &  98.8 & -30.1 / 26.8  & 0.7  &  \object{RECX 11}\\
J08475676-7854532 & 1.1e+30 & -2.92      & 12.0 & 2.46  &  97.2 & -30.4 / 26.3  & 1.2  &  \object{RECX 12}\\\hline
\multicolumn{7}{c}{new members}\\\hline\\[-3mm]
J08202975-8003259 & 3.0e+28 & -3.78 & 14.5  & 3.18 &  97.4 & -27.9 / 29.3 &  2.7 & core, RECX 19\\
J08575181-7738129 & 8.4e+28 & -3.51 & 13.4  & 2.59 &  93.8 & -33.3 / 26.1 &  5.2 & core, RECX 20\\
J09025131-7759347 & 2.0e+29 & -3.33 & 12.9  & 2.50 &  95.2 & -33.2 / 25.1 &  4.0 & core, RECX 21\\
J09053087-8134572 & 1.4e+28 & -3.73 & 15.7  & 3.54 & 100.0 & -32.1 / 23.4 &  5.1 & core, RECX 22\\
J08014860-8058052 & 1.2e+29 & -3.22 & 14.7  & 3.22 & 111.8 & -19.0 / 28.9 & 14.3 & ext., RECX 23\\\hline
\multicolumn{7}{c}{misc. pop.}\\\hline
J09353193-8018317 & 1.4e+29 & -3.12 & 14.3  & 2.54 & 116.9 & -23.2 / 23.8  &  19.2 & ext., iso7.4\\
J09424962-7726407 & 4.4e+29 & -3.08 & 12.8  & 2.02 & 111.9 & -25.9 / 22.9  &  14.9 & ext., iso7.4\\\hline
\end{tabular}
\end{center}
\tablefoot{$\eta_{\rm sep}$ denotes the 3D distance to the star $\eta$~Cha. Remarks: $G_{\rm phot}$ : Gaia EDR3 provides photometry only, search regions as defined in text, RECX member ID number as in Table \ref{mem}, and proposed designation for the new members.}
\end{table*}

To search for young, active stars, we combined the X-ray activity level $\log L_{\rm X}/L_{\rm bol}$ with the CMD position. For potential new members of the $\eta$~Cha cluster, parallax and proper motion values from Gaia EDR3 were taken into account, as discussed above. 
To support stellar youth, we required the stars to be reasonably elevated above the main sequence. We used the 100~Myr PARSEC isochrone as the dividing line. While the duration of the pre-main-sequence evolution depends on stellar mass, this is a suitable indicator to support stellar youth for our search of new members, which are expected to be rather late-type stars.
The very young stars are characterized by their proximity to the 10~Myr, 25~Myr, and 40~Myr isochrones. For potential $\eta$~Cha members, we required stars to be very young, that is, they needed to be located at or above the 10~Myr isochrone. In addition, we also inspected stars located around the 25~Myr isochrone because edge-on disk systems with gray extinction may appear below the isochrone. This is also the case for the members RECX~15 and 16, as shown above.

As with the astrometric and kinematic properties, all X-ray detected $\eta$~Cha members also fulfill the isochrone criterion and are classified as very young. Furthermore, out of the X-ray active field population, we identify five additional likely members of the $\eta$~Cha cluster. 

All member candidate stars (4 out of 4) found in the core search region are very young and are examined in detail further below.
In our extended region, we find another member candidate (1 out of 13) with an age of 10~Myr, but most of these stars are better described by the 25\,--\,100~Myr isochrones, and some are active stars already on the main sequence. 
Two stars in the extended search region have acceptable astrometric or kinematic and X-ray properties, but are better described by an older isochrone age and require further investigation. 
The majority of the more distant young stars at 140\,--\,200~pc also has isochrone ages of a several ten million years.
We briefly discuss other young populations located in the vicinity of or behind the cluster in Sect.\ref{dis}.

We compared our candidates with studies in the literature and searched for other activity, youth, or cluster member indicators.
Several of our candidates were also proposed as likely new $\eta$~Cha members by \cite[][M10]{mur10}. They searched a sky region of 5.5 deg radius around the cluster for young stars selected according to their position in IR CMDs. Their color selection criterion (i$_{\rm DENIS}$ - J$_{\rm 2MASS} > 1.5$) was specifically tailored to very red objects and was only supplemented by a few X-ray detections from the less sensitive RASS catalog. Using optical spectroscopy, they identified probable or possible members based on their kinematic properties, H$\alpha$ emission, and lithium-richness. 
Our eROSITA observation overlaps, but does not fully cover, their optical survey region. Notably, for all candidate stars that are investigated in both studies, the membership suggestion fully agrees.

Based on the X-ray data combined with the new Gaia measurements, we identified a set of $\eta$~Cha member candidates. These stars share all relevant properties with those of known $\eta$~Cha members, but are located farther away from the cluster core and exhibit a larger dispersion in their kinematic properties. We argue that the following stars very likely belong to the dispersed population of the cluster.

\smallskip
\noindent
\object{2MASS J08202975-8003259}. This star lies within the core search region and within 0.2 mag of the 10~Myr isochrone. It has a spectral type from Gaia BP-RP/2MASS H-K of M4.5/M4.5. It is a probable member in M10 with strong, variable H$\alpha$ emission.

\smallskip
\noindent
\object{2MASS J08575181-7738129}. This star lies within the core search region and within 0.2 mag of the 10~Myr isochrone. It has a spectral type from Gaia BP-RP/2MASS H-K of M3/M2.5. It lies beyond the blue selection limit in M10.

\smallskip
\noindent
\object{2MASS J09025131-7759347}. This star lies within the core search region and within 0.2 mag of the 10~Myr isochrone. It has a spectral type from Gaia BP-RP/2MASS H-K of M3/M2. It is a probable member in M10 (ROSAT candidate).

\smallskip
\noindent
\object{2MASS J09053087-8134572}. This star lies within the core search region and within 0.2 mag of the 10~Myr isochrone. It has a spectral type from Gaia BP-RP/2MASS H-K of M5.5/M4.5. It is a probable member in M10.

\smallskip
\noindent
\object{2MASS J08014860-8058052}. This star lies within the extended search region, 0.4 mag above 10~Myr isochrone, and is a potential binary. It has a spectral type from Gaia BP-RP/2MASS H-K of M4.5/M5.5. It is a possible member in M10.

The collective evidence appears to be overwhelming that these stars are physically associated with the cluster. We propose to add more memorable $\eta$~Cha cluster IDs in the same fashion as for the other members. The suggested associations are listed in Table~\ref{xmem}.

In addition to the new members, we find two stars that share many properties with the $\eta$~Cha cluster members, but belong to an apparently slightly older population with an age of about 25~Myr. They are located at a slightly larger distance of $\sim 115$~pc. Based on their kinematic properties, they are similar to the cluster members, but their origin is uncertain. Both are with $\log L_{\rm X}/L_{\rm bol} \approx -3$ highly active, comparable to the $\eta$~Cha members of similar spectral type. These high activity levels persist in the evolutionary process of M dwarfs over timescales of 100~Myr.

\subsection{Known $\eta$ Cha members}
\label{recx}

The known $\eta$~Cha members include three intermediate-mass systems with spectral types B8 to A7 and 15 late-type members with spectral types K6 to M5. No stars with spectral type F, G, or early K are among the members. Depending on the adopted mass estimates, a gap above or around one solar mass is present among the known members.

The star $\eta$~Cha (B8) itself and HD~75505 (A1) are undetected by eROSITA, similar as in other X-ray observations (see Sect. \ref{xhist}). This indicates that the X-ray detection of $\eta$~Cha (RECX~2) by ROSAT HRI was caused by UV contamination. A flaring very low-mass companion is unlikely as the HRI flux is classified as constant in \cite{mam00}.
With the exception of very young HAeBe stars and magnetic ApBp stars, late B- or early A-type stars are typically X-ray dark, and thus both stars may indeed be singles.

RS Cha is an eclipsing binary (A7+A8), but late A-type stars are also rather weak X-ray sources with cool 2\,--\,3~MK coronae \citep{rob09a}. Furthermore, its X-ray properties resemble those of other late-type members in the cluster. \cite{mam00} have proposed that RS~Cha might be a triple system. This hypothesis was further supported by the XMM data, in which the source showed a smaller flare in its light curve \citep{lop10}. In the eROSITA data, RS~Cha also exhibits an X-ray luminosity and coronal temperature distribution that are basically identical to those of early M-dwarf members of the cluster. 

We detect 14 out of 15 of the late-type members in X-rays. Only the faintest star 2MASS J08440914-7833457 (RECX~16) is below our sensitivity limit. However, ET~Cha (RECX~15) is missed by the detection pipeline and only appears as moderate photon excess ($20 \pm 10$ cts within 10\arcsec, adopted $f_{\rm X} = 2 \times 10^{-14}$\,erg\,cm$^{-2}$\,s$^{-1}$) in the PSF wing of EM~Cha, which is separated by about 40'' , but is nearly 100 times X-ray brighter. EN~Cha (RECX~9) lacks full Gaia information, and we adopted an ad hoc distance of 100~pc. EN~Cha is the only object that significantly falls below the X-ray level of $\log L_{\rm X}/L_{\rm bol} = -4$.

For all identified sources we provide 2MASS ID, Gaia photometric and kinematic information, as well as eROSITA X-ray properties in Table~\ref{xmem}. We extracted the X-ray information from the catalog data, that is bases on measured count rates and given fluxes refer to observed values. Therefore the X-ray luminosity of absorbed sources is formally a lower limit. In particular, the CTTS of spectral type M belong to the apparently least active stars in our sample. 
A dedicated spectral analysis is required to correct for these effects. With the exception of EN~Cha, a sufficient number of photons were collected to investigate variability and obtain useful spectra.

\begin{figure}[t]
\hspace{-0.5cm}\includegraphics[width=63mm,angle=-90]{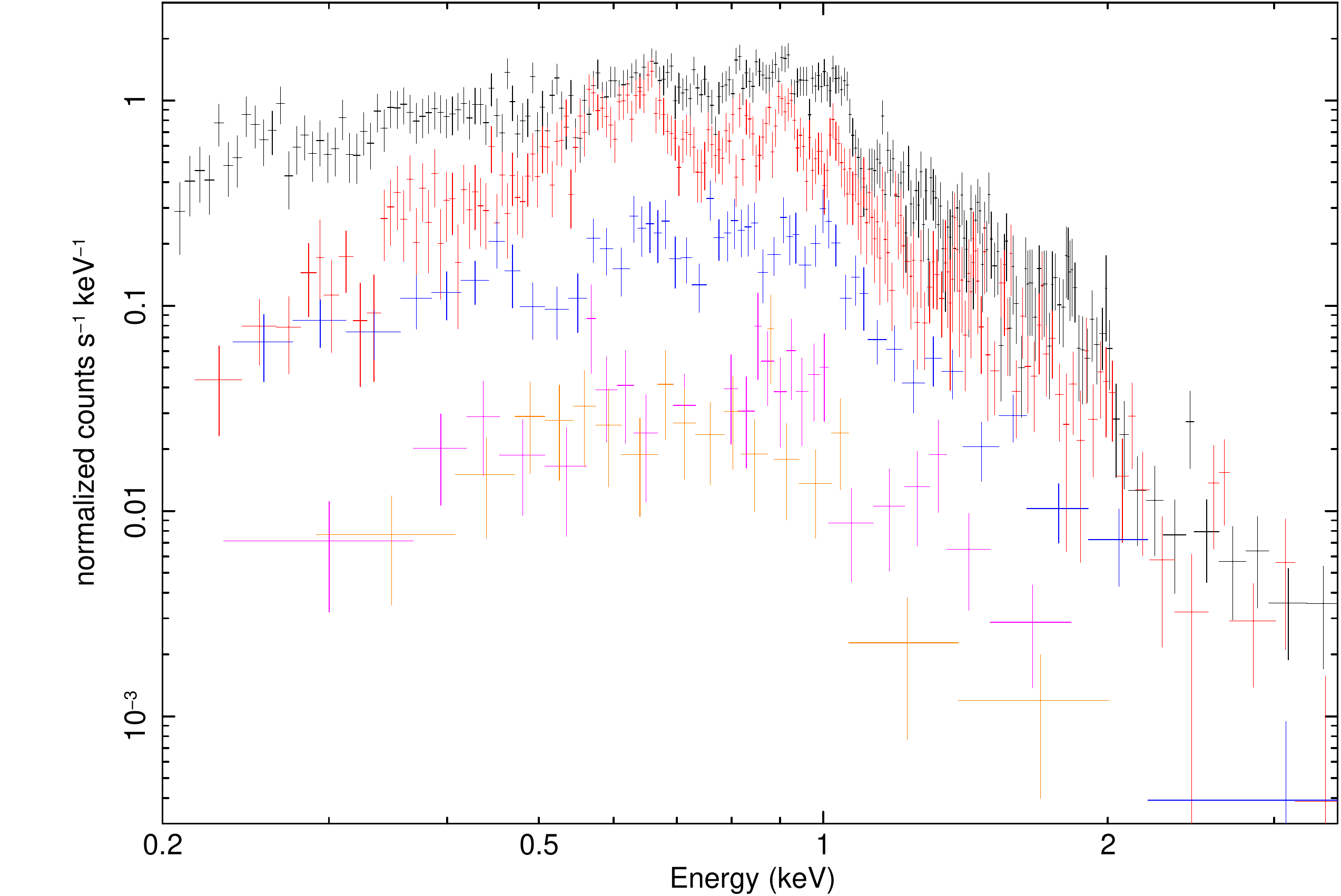}
\caption{\label{xsp} X-ray spectra of $\eta$~Cha members. We show the WTTS EG Cha (black) and EL Cha (blue), the TOs EH Cha (magenta) and EK Cha (orange), and the CTTS EP Cha (red). The spectra have a minimum of 5 counts/bin and are from the full eROSITA instrument.}
\end{figure}

\subsection{X-ray light curves and spectra}

\begin{table*}[t]
\caption{\label{sres}Spectral modeling results in the 0.2\,--\,5.0 keV band for RECX stars.}
\begin{center}
\renewcommand{\arraystretch}{1.2}
\begin{tabular}{crrrrrrrrr}\hline\hline\\[-4mm]
Star & $N_{\rm H}$ & $kT_1$ & $kT_2$ & $kT_3$ & $EM_1$ & $EM_2$ & $EM_3$ & $F_{\rm X\,obs}$ & $L_{\rm X\,emi}$\\
RECX & $10^{20}$ cm$^{-2}$ & keV & keV & keV & 10$^{52}$ cm$^{-2}$ & 10$^{52}$ cm$^{-3}$ & 10$^{52}$cm$^{-3}$ & erg\,cm$^{-2}$\,s$^{-1}$ & erg\,s$^{-1}$\\\hline
 1 & 1.42\,$^{+ 0.27 }_{- 0.28}$ & 0.25\,$^{+ 0.01 }_{- 0.01}$ & 0.85\,$^{+ 0.18 }_{- 0.11}$ & 1.25\,$^{+ 0.70 }_{- 0.07}$ & 11.60\,$^{+ 1.23 }_{- 1.31}$ & 6.98\,$^{+ 10.1}_{- 2.49}$ & 14.96\,$^{+ 2.64 }_{- 9.39}$ & 2.55e-12 & 3.4e+30\\
 3 & 4.44\,$^{+ 2.49 }_{- 2.85}$ & 0.19\,$^{+ 0.07 }_{- 0.08}$ & 0.87\,$^{+ 0.10 }_{- 0.06}$ & -- & 0.49\,$^{+ 1.92 }_{- 0.25}$ & 0.65\,$^{+ 0.15 }_{- 0.10}$ & -- & 6.74e-14 & 1.1e+29\\
 4 & 2.61\,$^{+ 0.57 }_{- 0.55}$ & 0.26\,$^{+ 0.02 }_{- 0.02}$ & 1.01\,$^{+ 0.04 }_{- 0.03}$ & -- & 4.25\,$^{+ 0.61 }_{- 0.58}$ & 5.00\,$^{+ 0.30 }_{- 0.30}$ & -- & 6.33e-13 & 9.3e+29\\
 5 & 3.38\,$^{+ 6.10 }_{- 3.13}$ & 0.17\,$^{+ 0.04 }_{- 0.05}$ & 0.72\,$^{+ 0.05 }_{- 0.16}$ & -- & 0.47\,$^{+ 1.01 }_{- 0.20}$ & 0.33\,$^{+ 0.08 }_{- 0.07}$ & -- & 4.28e-14 & 7.2e+28\\
 6 & 2.61\,$^{+ 0.83 }_{- 0.77}$ & 0.30\,$^{+ 0.03 }_{- 0.03}$ & 1.11\,$^{+ 0.07 }_{- 0.06}$ & -- & 2.34\,$^{+ 0.37 }_{- 0.35}$ & 2.58\,$^{+ 0.24 }_{- 0.23}$ & -- & 3.42e-13 & 5.0e+29\\
 7 & 1.61\,$^{+ 0.51 }_{- 0.51}$ & 0.26\,$^{+ 0.03 }_{- 0.03}$ & 0.82\,$^{+ 0.15 }_{- 0.13}$ & 1.27\,$^{+ 0.79 }_{- 0.14}$ & 7.63\,$^{+ 1.65 }_{- 1.59}$ & 7.96\,$^{+ 4.59 }_{- 3.05}$ & 7.46\,$^{+ 3.25 }_{- 5.20}$ & 1.77e-12 & 2.4e+30\\
 8 & 2.35\,$^{+ 1.05 }_{- 0.49}$ & 0.34\,$^{+ 0.02 }_{- 0.04}$ & 1.02\,$^{+ 0.07 }_{- 0.07}$ & -- & 3.65\,$^{+ 0.77 }_{- 0.72}$ & 3.13\,$^{+ 0.46 }_{- 0.23}$ & -- & 4.82e-13 & 6.9e+29\\
 10 & 2.09\,$^{+ 1.22 }_{- 0.91}$ & 0.14\,$^{+ 0.06 }_{- 0.03}$ & 0.35\,$^{+ 0.05 }_{- 0.03}$ & 1.15\,$^{+ 0.07 }_{- 0.09}$ & 1.11\,$^{+ 1.30 }_{- 0.69}$ & 2.48\,$^{+ 0.40 }_{- 0.58}$ & 2.83\,$^{+ 0.25 }_{- 0.26}$ & 4.27e-13 & 6.1e+29\\
 11 & 9.84\,$^{+ 0.89 }_{- 0.77}$ & 0.22\,$^{+ 0.01 }_{- 0.01}$ & 0.86\,$^{+ 0.05 }_{- 0.02}$ & 1.29\,$^{+ 0.18 }_{- 0.09}$ & 18.04\,$^{+ 1.77 }_{- 1.54}$ & 4.29\,$^{+ 0.41 }_{- 0.43}$ & 6.25\,$^{+ 1.12 }_{- 1.09}$ & 1.10e-12 & 2.6e+30\\
 12 & 1.36\,$^{+ 0.38 }_{- 0.37}$ & 0.28\,$^{+ 0.02 }_{- 0.02}$ & 0.95\,$^{+ 0.12 }_{- 0.18}$ & 1.53\,$^{+ 0.37 }_{- 0.18}$ & 4.59\,$^{+ 0.66 }_{- 0.77}$ & 2.88\,$^{+ 1.77 }_{- 1.29}$ & 5.81\,$^{+ 1.35 }_{- 2.03}$ & 1.05e-12 & 1.3e+30\\
 14 & 3.45\,$^{+ 14.3}_{- 3.45}$ & 1.29\,$^{+ 0.33 }_{- 0.24}$ & -- & -- & 0.27\,$^{+ 0.12 }_{- 0.08}$ & -- & -- & 1.99e-14 & 2.8e+28\\
 17 & 0.00\,$^{+ 2.10 }_{- 0.00}$ & 0.20\,$^{+ 0.06 }_{- 0.06}$ & 0.73\,$^{+ 0.22 }_{- 0.19}$ & -- & 0.34\,$^{+ 0.16 }_{- 0.08}$ & 0.21\,$^{+ 0.09 }_{- 0.10}$ & -- & 4.28e-14 & 5.1e+28\\
 18 & 1.0 & 0.27\,$^{+ 0.04 }_{- 0.03}$ & -- & -- & 0.26\,$^{+ 0.06 }_{- 0.05}$ & -- & -- & 1.67e-14 & 2.4e+28\\\hline

\multicolumn{10}{l}{new members}\\
{\small 080148-8058} & 18.56\,$^{+ 9.64 }_{- 7.26}$ & 0.28\,$^{+ 0.08 }_{- 0.05}$ & 1.30\,$^{+ 0.26 }_{- 0.16}$ & -- & 1.67\,$^{+ 1.90 }_{- 0.86}$ & 1.42\,$^{+ 0.31 }_{- 0.29}$ & -- & 8.33e-14 & 3.0e+29\\
{\small 082029-8003} & 0.00\,$^{+ 2.01 }_{- 0.00}$ & 0.34\,$^{+ 0.13 }_{- 0.08}$ & 1.64\,$^{+ 1.60 }_{- 0.42}$ & -- & 0.14\,$^{+ 0.08 }_{- 0.03}$ & 0.22\,$^{+ 0.09 }_{- 0.08}$ & -- & 3.14e-14 & 3.6e+28\\
{\small 085751-7738} & 2.24\,$^{+ 2.49 }_{- 1.97}$ & 0.22\,$^{+ 0.11 }_{- 0.05}$ & 0.81\,$^{+ 0.17 }_{- 0.07}$ & -- & 0.33\,$^{+ 0.24 }_{- 0.15}$ & 0.64\,$^{+ 0.09 }_{- 0.15}$ & -- & 7.90e-14 & 1.0e+29\\
{\small 090251-7759} & 2.05\,$^{+ 1.12 }_{- 1.08}$ & 0.23\,$^{+ 0.03 }_{- 0.02}$ & 1.02\,$^{+ 0.06 }_{- 0.05}$ & -- & 1.02\,$^{+ 0.29 }_{- 0.27}$ & 1.54\,$^{+ 0.15 }_{- 0.14}$ & -- & 1.97e-13 & 2.6e+29\\
{\small 090530-8134} & 0.00\,$^{+ 2.03 }_{- 0.00}$ & 0.31\,$^{+ 0.11 }_{- 0.05}$ & -- & -- & 0.17\,$^{+ 0.05 }_{- 0.04}$ & -- & -- & 1.34e-14 & 1.6e+28\\\hline
\end{tabular}
\end{center}
\end{table*}

The X-ray light curves obtained in a field-scan mode observation have a temporal sampling that naturally follows the source coverage pattern, but further depends on the position of the object within the field. As an example for a source located close to the field center, the light curve for EG~Cha (aka RECX~1) is shown in in Fig.~\ref{xlc}. The light curve is binned to 100~s and is vignetting corrected. Bins with a fractional exposure below 0.2 were discarded. Ten field-of-view passages, each lasting between 100 and 500 seconds, are distributed in total over about 8 hours. Very similar patterns track the temporal evolution of the other stellar sources at the cluster center. As an example for a source located in a corner of the observed field, a flaring mid-M dwarf is shown in the bottom panel of the same figure. The star is located at 90~pc distance and very active, but it is not associated with the $\eta$~Cha cluster based on its estimated age of about 50~Myr and its proper motion vector.

The light curve of EG~Cha and the light curves of the other studied $\eta$~Cha members are overall quite flat. We detect only mild X-ray variability at a level of about 20\,\% or even lower in the available data. No strong flaring occurred during the observation in any of our investigated targets, and thus the determined properties reflect the quasi-quiescent state of these stars.

We extracted spectra for all targets of interest for which sufficient source photons were detected by eROSITA.
Photon numbers range between roughly 30 for the faintest targets and 6000 for the brightest ones, and the spectral quality differs accordingly.
Example spectra obtained by eROSITA during the observation are shown in Fig.~\ref{xsp} for several YSOs of various spectral types and evolutionary states. Spectra are typically more highly absorbed for younger, that is, disk-bearing stars. The evolutionary sequence CTTS - TO - WTTS is also visible in the X-ray spectra. To quantify the coronal properties of the cluster members, we performed a spectral modeling of all X-ray brighter members for which sufficient counts are available in the 0.2\,--\,5.0~keV range.
The different quality of the spectra was taken into account by varying the complexity of the respective applied spectral model.
Our modeling results are given in Table~\ref{sres}.

\begin{figure*}[t]
\includegraphics[width=180mm]{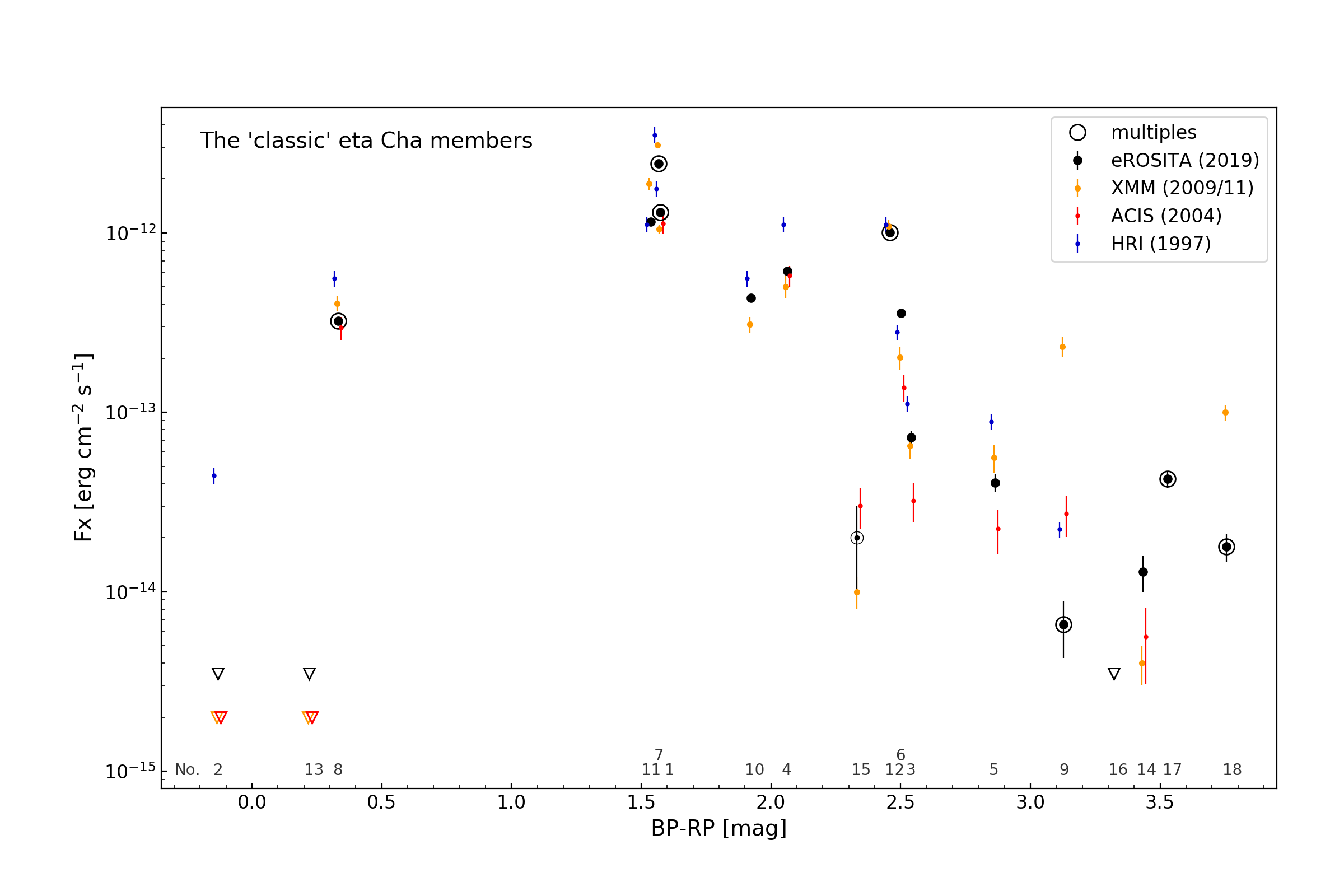}
\caption{\label{xray_bprp}X-ray history of $\eta$~Cha cluster observations. Detected X-ray flux vs. Gaia BP-RP color for the established cluster members from various missions with upper limits plotted as downward triangles. The $\eta$~Cha RECX number is given in the bottom of the plot.}
\end{figure*}

The determined plasma properties indicate that the X-ray emission of the $\eta$~Cha member is predominantly created by magnetic activity.
The typically broad temperature distribution and the presence of plasma at or above 10~MK in basically all studied stars indicates a coronal origin of the detected X-ray emission. This is expected at the cluster age of about 10~Myr, and many of the X-ray bright stars are indeed classified as WTTS.
In the stars that are classified as disk-bearing, that is, which have SED class II, flat, TO, strong H$\alpha$ emission as sign of ongoing accretion is observed only in a few late-type members (RECX 5, 15, and 16). However, most of them are too faint in X-rays to derive firm conclusions about potential X-ray emission from accretion shocks with our data. The CTTS EP Cha (RECX~11) stands out in X-rays with a strong cool component, a high EM1-to-EM2 ratio, and high absorption column, but it did not show strong accretion signatures like $H\alpha$ in previous optical studies.

Four out of five of the new members show plain WTTS characteristics, that is, very low absorption and an active corona. An exception is  
2MASS J08014860-8058052, where spectral modeling indicates significant absorption, likely due to the presence of a circumstellar disk,
with the caveat that the error on its $N_{\rm H}$ is quite substantial.

\subsection{Long-term variability and cross-check with other missions}
\label{xhist}

The $\eta$~Cha cluster has been repeatedly observed in X-rays in the past decades.
This X-ray history enables a comparison of our results and to test the long-term variability of the cluster members that are covered by the respective missions. We adopted results from \cite{mam99} for ROSAT/HRI and from \cite{lop10} for XMM/2009. We determined X-ray fluxes from our own analysis of the archival data when no literature values were available. For the {\it Chandra} ACIS-I observation (OBSID 4491, 10~ks), we derived fluxes from the measured count rates, and using an ECF of $7.0 \times 10^{-12}$~erg\,s$^{-1}$ per cts\,s$^{-1}$ , we calculated with {\it Chandra} PIMMS. For the {\it XMM-Newton} observation (OBSID 0652330101, 40~ks) of the region around EG~Cha taken in 2011, we processed the dataset with the SAS software package and performed a spectral analysis of the covered cluster members, here EG~Cha (RECX~1) and RECX~18.

A comparison of the X-ray brightness of $\eta$~Cha members obtained by various missions in about two decades is shown in Fig.~\ref{xray_bprp}. Corrections for slight differences in the energy band between the instruments and source distance between the stars are at a level of 10\,\% and are neglected here.

The derived fluxes do not show any obvious trends between the instruments, except for ROSAT/HRI, which tends to give about 50\,\% higher fluxes. Source fluxes typically differ mildly between the observations, and variations by about 30\,\% up to a factor of about two are the most commonly observed variability level. When much stronger variability is present, it is commonly associated with clear flaring events. RECX~9 was observed during a larger flare with XMM, and RECX~18 also shows flaring activity in the XMM observations. The flux differences are of factor 35 and 5 when compared to eROSITA. RECX~9 was particularly faint in the eROSITA observation, even about a factor four fainter than in the {\it Chandra} observation.

\begin{figure}[t]
\hspace{-0.2cm}\includegraphics[width=60mm,angle=-90]{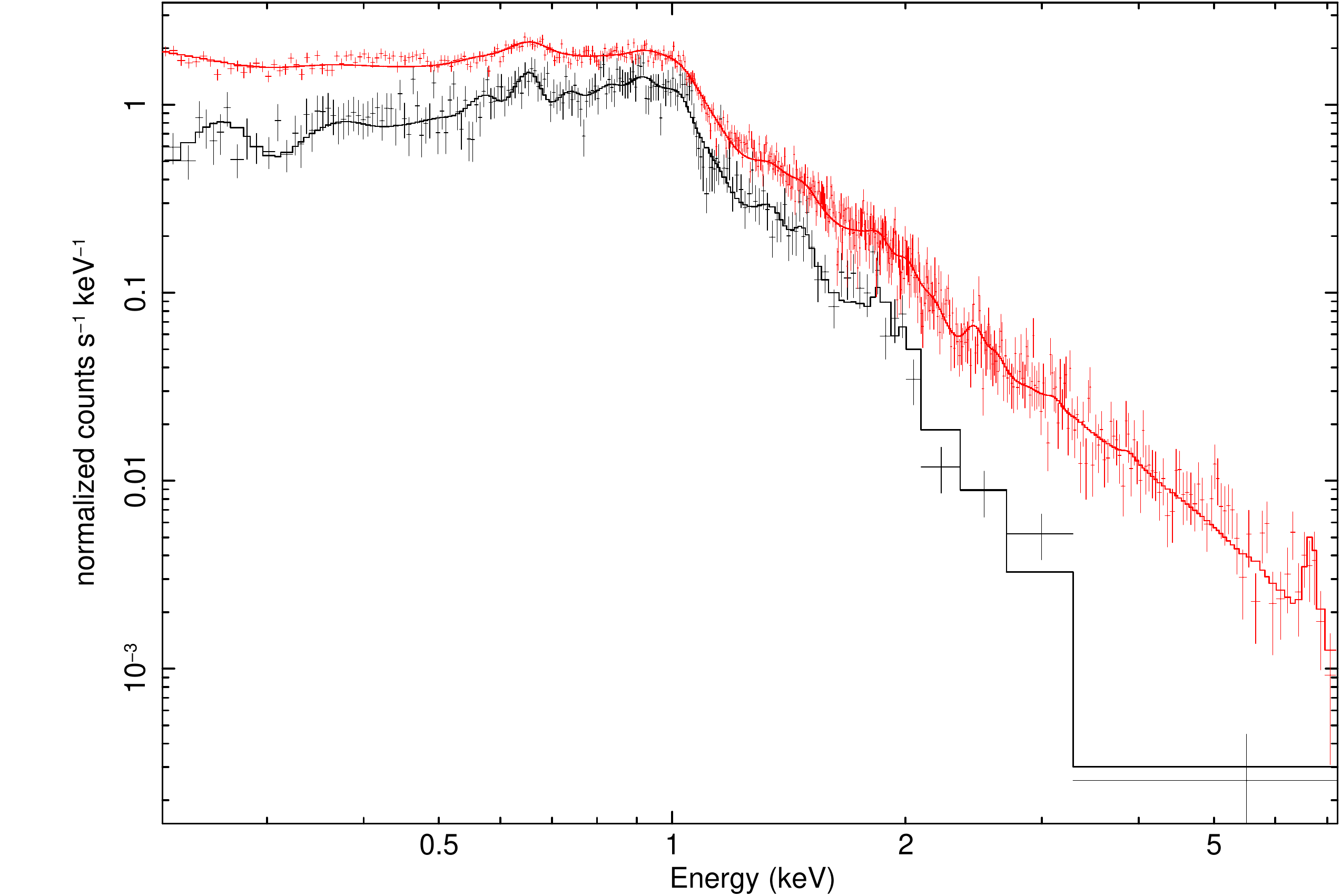}

\vspace*{-3.1cm}
\hspace*{1.cm}
\includegraphics[width=40mm]{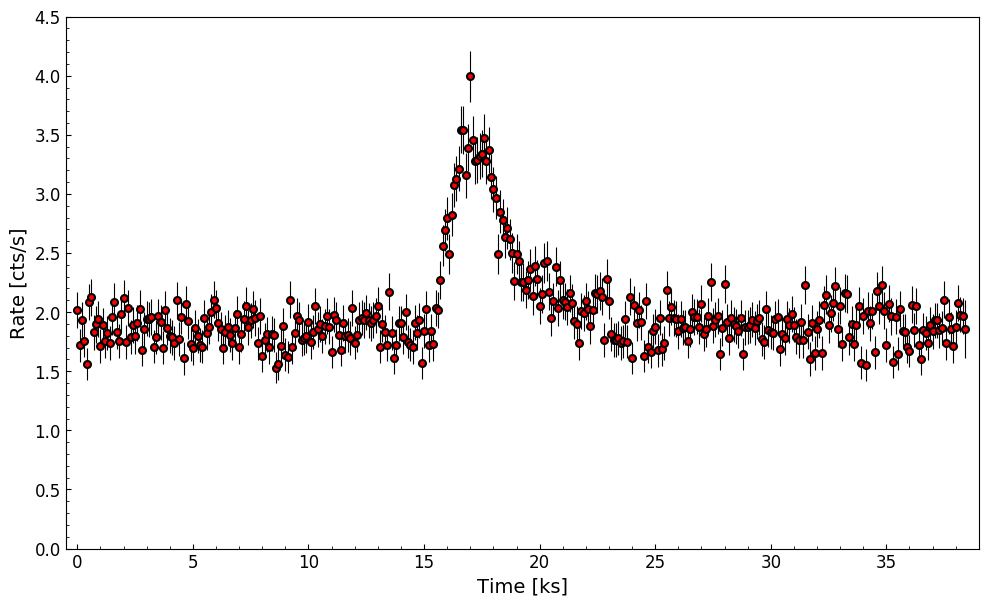}
\vspace*{0.5cm}
\caption{\label{xsp2} X-ray spectra of EG Cha obtained by eROSITA (black) and XMM pn (red) and resp. spectral model. The XMM light curve (inset), shows a moderate flare within the observations.}
\end{figure}

\begin{table}[t]
\begin{center}
\caption{\label{egchaspec}EG Cha. Spectral comparison of eROSITA vs. XMM/pn.}
\renewcommand{\arraystretch}{1.2}
\setlength{\tabcolsep}{2pt}
\begin{tabular}{lrrrr}\hline\hline\\[-4mm]
 & eROSITA & XMM  & eROSITA & XMM  \\\hline
$N_{\rm H}$ [10$^{20}$ cm$^{-2}$] & {1.56\,$^{+ 0.27}_{- 0.27}$}  & {1.53\,$^{+ 0.13}_{- 0.12}$} & \multicolumn{2}{c}{1.55\,$^{+ 0.11}_{- 0.11}$} \\
$kT_1$ [keV] & 0.26\,$^{+ 0.01}_{- 0.01}$ & 0.26\,$^{+ 0.01}_{- 0.01}$ & \multicolumn{2}{c}{0.26\,$^{+ 0.01}_{- 0.01}$} \\
$kT_2$ [keV] & 1.01\,$^{+ 0.26}_{- 0.26}$ & 1.05\,$^{+ 0.01}_{- 0.01}$ & \multicolumn{2}{c}{1.04\,$^{+ 0.01}_{- 0.01}$} \\
$kT_3$ [keV] & 1.82\,$^{+ 1.07}_{- 0.73}$ & 2.81\,$^{+ 0.18}_{- 0.15}$ & \multicolumn{2}{c}{2.77\,$^{+ 0.15}_{- 0.17}$} \\
$EM_1$ [10$^{52}$ cm$^{-3}$] & 12.2\,$^{+ 1.2}_{- 1.2}$  & 15.5\,$^{+ 0.5}_{- 0.5}$  & 12.7\,$^{+ 0.7}_{- 0.7}$  & 15.5\,$^{+ 0.4}_{- 0.4}$ \\
$EM_2$ [10$^{52}$ cm$^{-3}$]& 16.4\,$^{+ 1.4}_{- 1.3}$  & 19.1\,$^{+ 0.4}_{- 0.4}$ & 17.7\,$^{+ 0.8}_{- 0.8}$  & 18.9\,$^{+ 0.4}_{- 0.5}$ \\
$EM_3$ [10$^{52}$ cm$^{-3}$]&  4.0\,$^{+ 0.5}_{- 0.5}$  &  9.9\,$^{+ 0.6}_{- 0.6}$&  2.3\,$^{+ 0.9}_{- 0.9}$  & 10.1\,$^{+ 0.6}_{- 0.5}$  \\\hline
$\chi^2$ (d.o.f.) & \multicolumn{2}{c}{1.20 (705)}  & \multicolumn{2}{c}{1.19 (709)} \\\hline
$L_{\rm X~obs}$ [10$^{30}$ erg\,s$^{-1}$] & 2.7/3.6 & 2.9/4.1 & 2.7/3.6 & 2.9/4.0 \\
$L_{\rm X~emi}$ [10$^{30}$ erg\,s$^{-1}$] & 3.1/4.2 & 3.3/4.6 & 3.1/4.2 & 3.3/4.6 \\\hline
\end{tabular}
\tablefoot{$L_{\rm X}$ refers to 0.2\,--\,2.3  and 0.2\,--\,5.0 keV band.}
\end{center}
\end{table}

A dedicated spectral comparison of data taken by eROSITA and {\it XMM-Newton} was performed for EG~Cha. We extracted EPIC spectra and modeled them together with the eROSITA data in an identical manner as above, but used a binning with a minimum of 15 counts. A comparison of the spectra taken by eROSITA and XMM pn is shown in Fig.~\ref{xsp2}. The XMM observation includes a moderate flare, which is shown as an inset. Because EG Cha was found to be slightly brighter than during the eROSITA observation even outside the flare and normalizations are mainly affected, we used all available data in the analysis.

Both spectra were fit first with a set of independent models and then with a combined model, in which only the normalizations of the spectral components were free parameters. The fit results are given in Table~\ref{egchaspec}. Only minor differences and a basically identical fit quality are present. The determined absorption column as well as plasma temperatures agree within the errors between all modeling approaches. Considering that the pn spectrum contains about ten times more photons, the accuracy of the fit results obtained from the eROSITA data is excellent. 
The preference of the eROSITA model for a lower temperature of the hottest plasma component is notable. This is similar when we compare our results for other studied stars with the literature values. However, these model components are less constrained as the eROSITA spectrum drops in photon number comparably more strongly above 2.3~keV because the effective area of the instrument at higher energies is smaller.

\begin{figure*}[t]
\includegraphics[trim={3.cm 3.cm 3.cm 3.cm}, clip, width=180mm]{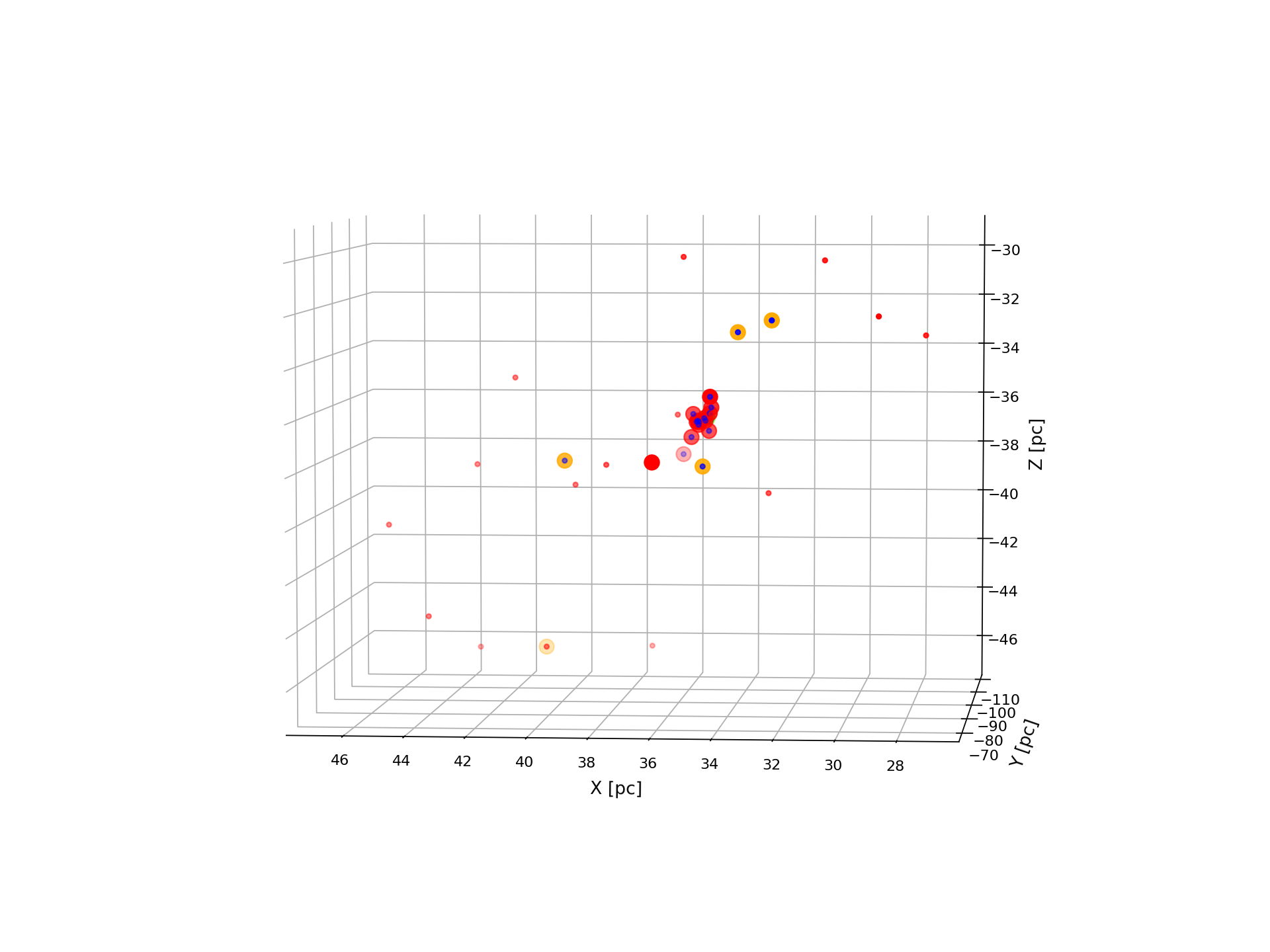}
\caption{\label{gview}3D galactic view of the $\eta$~Cha cluster and its surrounding halo of young stars. The Sun is located at (0,0,0), the x-axis is directed toward the Galactic center, the y-axis in the direction of galactic rotation, and the z-axis lies perpendicular to the Galactic plane. Large circles indicate known (red) and new (yellow) members, small circles denote stars found in our core (blue) and extended search region (red), as defined in Sect.~\ref{ids}.}
\end{figure*}

\begin{figure}[ht]
\includegraphics[width=92mm]{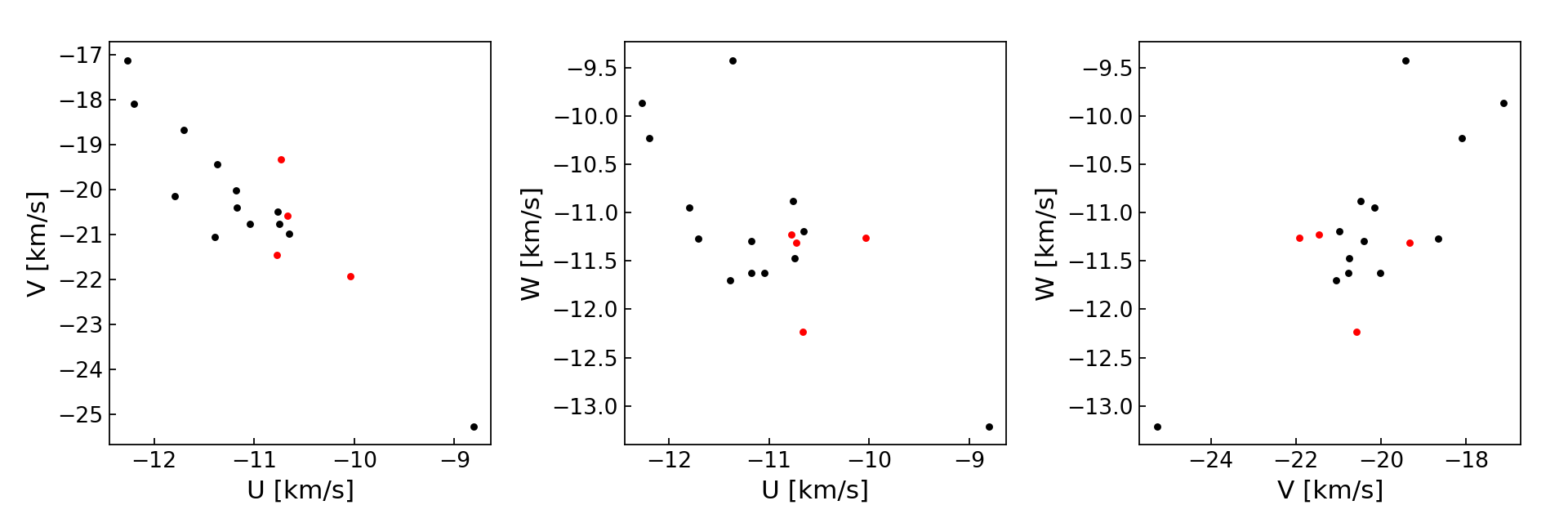}
\caption{\label{uvw}Galactic (U, V, W) velocities  of know (black) and new (red) $\eta$~Cha members.}
\end{figure}

\section{Discussion}
\label{dis}

The eROSITA PV observation of the $\eta$~Cha cluster has been used to identify ejected low-mass members through their X-ray activity and Gaia data. With the five new members, the $\eta$~Cha cluster population increases to at least 23. The new members are mid-M dwarfs of spectral types of about M3 to M5 and have masses of about $0.1 - 0.3~M_{\odot}$. An extended region around $\eta$~Cha has been surveyed about a decade ago in the optical by \cite{mur10}. When stars are covered in both studies, membership was independently confirmed by our eROSITA/Gaia analysis and their work.

Depending on the respective masses and ages of the studied stellar population, specific indicators may be more robust, but the various combinations of kinematic properties and different youth indicators such as CMD position, X-ray activity, or lithium-richness complement each other very well in the study of nearby young stellar populations. In this respect, the eROSITA all-sky survey brings the available X-ray data to a new level.

The Gaia data allow us to construct a 3D view of the cluster and the surrounding X-ray emitting stars in Galactic coordinates, as shown in Fig.~\ref{gview}. The eROSITA observation is quite sensitive to potential members in both radial directions, but it is limited in transversal direction by the field boundaries.
At the distance of the cluster, an angular separation of $1^{\circ}$ corresponds to about 1.7~pc distance in the plane of the sky. Therefore only sources located within a radius of about 4.5~pc around the cluster center are surveyed completely by the eROSITA observation. As we detect two members at distances beyond 5~pc and one at 15~pc, a large fraction of the potential member halo might not be covered by our data. 

Assuming a spherical distribution of the ejected stellar population, another 10\,--\,20 members, but with significant error, can be expected in a comparable sensitive X-ray observation beyond the surveyed region. Furthermore, some later M dwarfs may be emitting below our detection limit and could have been missed. Potential cluster members that lie in the regime of very low-mass stars or brown dwarfs, that is, with masses of $M < 0.1~M_{\odot}$, are too faint and remain undetected in our study.

However, the starting conditions of the cluster are ambiguous. In contrast to the N-body simulations by \cite{mor07}, where a compact high-density configuration was the favored initial state, other scenarios have been suggested. \cite{beck13} also used N-body simulations and favored an initial state that was basically devoid of very low-mass objects and wide binaries due to particular initial conditions. They suggested that dynamical evolution did not play a major role. Still, ejected members and tighter binaries with system masses above 0.1~$M_{\odot}$ are at least sparsely present in a halo population around the cluster.

We also inspected the kinematics of our targets in Galactic heliocentric velocities U, V, and W. Radial velocities from Gaia are only available for a few of our stars, for example, 2 out of 18 for the known members. When available, we additionally used RV values for the known members from \cite{gag18}, \cite{schn19} and the Simbad database. For the new members, the values are from \cite{mur10}. The result is shown in Fig.~\ref{uvw}, but we caution that the RV errors are quite large for some stars. The proposed new members are found to be kinematically well within the dispersion of the known members, suggesting a basically comoving group. We observe some scatter, however, and as a velocity of 1 km\,s$^{-1}$ corresponds to a travel distance of about 1 pc\,Myr$^{-1}$, stellar displacements of a few up to 15\,pc are easily reached within the cluster lifetime even for moderate ejection speeds of a few km\,s$^{-1}$. 

In addition to the cluster members, we find many other young stars that are located around the $\eta$~Cha cluster. These might even have similar astrometric and kinematic properties, but are more widespread and appear to originate from a slightly older 20\,--\,50 Myr population. We tested the active stars found in the extended search region with the BANYAN $\Sigma$ tool \citep{gag18} for membership of young stellar associations. We find, if any, as the most likely origin dispersed members of the Lower Centaurus-Crux (LCC) population. Behind the $\eta$~Cha cluster, several more or less pronounced peaks in the number distribution at 140, 160, and 180~pc are present. These distances correspond to the known distances of the Chamaeleon I and II clouds, and although the detected stars were not found to be exceptionally young, previous star formation episodes that have occurred in the vicinity of these regions are potential origins.

\section{Summary}
\label{sum}

We have observed the $\eta$~Cha cluster with eROSITA and derived the X-ray properties of known members. We also identify likely new members. Our main findings are listed below.

\begin{enumerate}
\item The B8V star $\eta$~Cha remained undetected. The A-type binary RS~Cha exhibits X-ray properties similar to early-M~dwarfs in the cluster. None of the three more massive components is likely an intrinsic X-ray emitter.

\item We detected 14 out of 15 of the known low-mass members of the cluster. Only the faintest target remained undetected. We achieved a detection completeness in the range of 90\,\% for stars above $0.1 M_{\odot}$ . Their X-ray emission is of coronal origin.

\item Five new members of the $\eta$~Cha cluster are found in our eROSITA and Gaia study based on X-ray activity and astrometric and photometric properties. Several of these stars also show other youth indicators such as lithium richness and were identified as likely members by \cite{mur10}.

\item The total number of $\eta$~Cha members is now at least 23 and has increased by a factor of about two since the initial study. Further low-mass members that were ejected early in the cluster history and/or that have high velocity are expected to be located beyond the surveyed region, however.
\end{enumerate}

\begin{acknowledgements}
This work is based on data from eROSITA, the soft X-ray instrument aboard SRG, a joint Russian-German science mission supported by the Russian Space Agency (Roskosmos), in the interests of the Russian Academy of Sciences represented by its Space Research Institute (IKI), and the Deutsches Zentrum f{\"u}r Luft- und Raumfahrt (DLR). The SRG spacecraft was built by Lavochkin Association (NPOL) and its subcontractors, and is operated by NPOL with support from the Max Planck Institute for Extraterrestrial Physics (MPE).

The development and construction of the eROSITA X-ray instrument was led by MPE, with contributions from the Dr. Karl Remeis Observatory Bamberg \& ECAP (FAU Erlangen-Nuernberg), the University of Hamburg Observatory, the Leibniz Institute for Astrophysics Potsdam (AIP), and the Institute for Astronomy and Astrophysics of the University of Tuebingen, with the support of DLR and the Max Planck Society. The Argelander Institute for Astronomy of the University of Bonn and the Ludwig Maximilians Universitaet Munich also participated in the science preparation for eROSITA. The eROSITA data shown here were processed using the eSASS software system developed by the German eROSITA consortium.
J.R. acknowledges support from the DLR under grant 50QR2105.
This research has made use of the SIMBAD database, operated at CDS, Strasbourg, France.
We thank the referee, Eric E. Mamajek, for constructive comments.

\end{acknowledgements}

\bibliographystyle{aa}
\bibliography{etacha}

\begin{appendix}

\section{Description of catalog entries}
\label{app}

\begin{table}[ht]
\caption{X-ray source catalog of the eROSITA $\eta$ Cha field scan.}
\begin{tabular}{lrrr}\hline\\[-3mm]
Name         & Format & Unit & Description \\\hline\\[-3mm]
    \multicolumn{4}{l}{Columns in the main catalog (0.2\,--\,2.3~keV)}\\\hline\\[-3mm]
ERO\_NAME       & 21A & ...  & eROSITA/SRG name\\
ID\_SRC         & J   & ...  & Source ID\\
RA              & D   & deg & Right ascension\\
DEC             & D   & deg & Declination\\
RADEC\_ERR      & E   & arcsec  & Combined positional error\\
EXT             & E   & arcsec  & Source extent\\
EXT\_ERR        & E   & arcsec  & Extent error\\
EXT\_LIKE       & E   & ... & Extent likelihood\\
ML\_CTS         & E   & cts & Source counts from PSF-fitting\\
ML\_CTS\_ERR    & E   & cts & 1-$\sigma$ source counts error\\
ML\_RATE        & E   & cts\,s$^{-1}$ & Source rate from PSF-fitting\\
ML\_RATE\_ERR   & E   & cts\,s$^{-1}$ & 1-$\sigma$ rate error\\
ML\_FLUX        & E   & erg\,cm$^{-2}$\,s$^{-1}$ & Source flux in the detection band \\
ML\_FLUX\_ERR   & E   & erg\,cm$^{-2}$\,s$^{-1}$ & 1-$\sigma$ source flux error\\
DET\_LIKE       & E   & ... & Detection likelihood\\
ML\_BKG         & E   & cts\,arcmin$^{-2}$ & Background at the source position\\\hline\\[-3mm]

\multicolumn{4}{l}{Columns in the hard catalog (2.3\,--\,5.0~keV)}\\\hline\\[-3mm]
\multicolumn{4}{l}{as above, except ERO\_NAME replaced by:}\\

ID\_MAIN &J &... & Source ID in the main catalog for overlapping sources\\
\hline  
\end{tabular}
\end{table}

\end{appendix}

\end{document}